\documentclass[prl,superscriptaddress,showpacs,longbibliography,reprint]{revtex4-2} 

\usepackage[colorlinks=true,linkcolor=blue,urlcolor=blue,citecolor=blue,pdfusetitle]{hyperref}
\usepackage{color}
\usepackage[utf8]{inputenc}
\usepackage[usenames,dvipsnames]{xcolor}
\usepackage{amsthm}
\usepackage{amsmath}
\usepackage{amssymb}
\usepackage{graphicx}
\graphicspath{{graphics/}}
\usepackage{epsfig}
\usepackage{dcolumn}% Align table columns on decimal point
\usepackage{bm}% bold math
\usepackage{mathrsfs}
\usepackage{multirow}
\usepackage[all]{xy}
\usepackage{pbox}
\usepackage{lipsum}
\usepackage{verbatim}
\usepackage{braket}
\usepackage{dsfont}
\usepackage{array}
\usepackage{makecell}
\usepackage{tabularx}
\usepackage{isotope}
\usepackage{xr}
\usepackage{float}

\newtheorem{theorem}{Theorem}
\newtheorem{lemma}{Lemma}
\newtheorem{proposition}{Proposition}

\newcommand{\tr}{\mathrm{tr}}

\begin{document}
\raggedbottom

\title{Long-range interacting systems are locally non-interacting}

\newcommand{\tubingen}{Institut f\"{u}r Theoretische Physik and Center for Integrated Quantum Science and Technology,  Universit\"{a}t T\"{u}bingen, Auf der Morgenstelle 14, 72076 T\"{u}bingen, Germany}

\author{Robert Mattes}
\affiliation{\tubingen}

\author{Igor Lesanovsky}
\affiliation{\tubingen}
\affiliation{School of Physics and Astronomy and Centre for the Mathematics and Theoretical Physics of Quantum Non-Equilibrium Systems, The University of Nottingham, Nottingham, NG7 2RD, United Kingdom}

\author{Federico Carollo}
\affiliation{Centre for Fluid and Complex Systems, Coventry University, Coventry, CV1 2TT, United Kingdom}

\date{\today}

\begin{abstract}
Enhanced experimental capabilities to control nonlocal and power-law decaying interactions are currently fuelling intense research in the domain of quantum many-body physics. 
Compared to their counterparts with short-ranged interactions, long-range interacting systems display novel physics, such as nonlinear light cones for the propagation of information or inequivalent thermodynamic ensembles. In this work, we consider generic long-range open quantum systems in arbitrary dimensions and focus on the so-called {\it strong long-range} regime. We prove that in the thermodynamic limit local properties, captured by reduced quantum states, are described by an emergent non-interacting theory. Here, the dynamics factorizes and the individual constituents of the system evolve independently such that no correlations are generated over time. In this sense, long-range interacting systems are locally non-interacting. This has significant implications for their relaxation behavior, for instance in relation to the emergence of long-lived quasi-stationary states or to the absence of thermalization.
\end{abstract}

\maketitle

\begin{figure}[t!]
    \centering
    \includegraphics[width=\columnwidth]{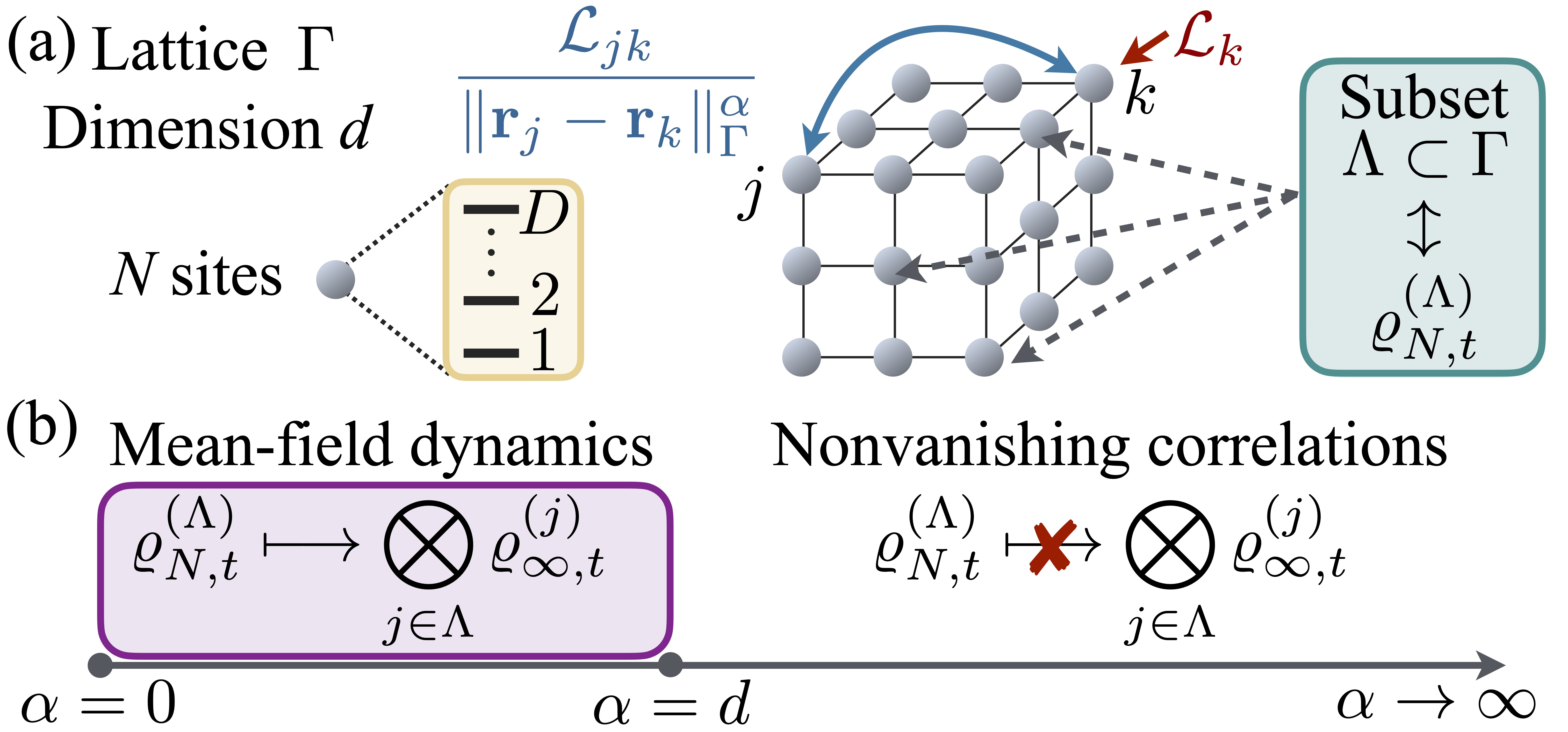}
    \caption{{\bf Long-range interacting quantum systems.} (a) Quantum system consisting of $N$ $D$-level particles, arranged on a $d$-dimensional lattice $\Gamma$. The vector $\textbf{r}_k$ gives the position of the $k$th lattice site. The local dynamics is described by the generator  $\mathcal{L}_k$. The two-body long-range interactions are encoded in $\mathcal{L}_{jk}$ and their strength decays as the inverse power (exponent $\alpha$) of the interparticle distance. 
    (b) For $\alpha\leq d$ the reduced state $\varrho_{N,t}^{(\Lambda)}$, associated with a subset $\Lambda$ of sites [see panel (a)], remains in product form, in the thermodynamic limit, if initially a product state. This is a characteristic feature of mean-field theories. On the contrary, for $\alpha > d$, nonvanishing correlations can be, in general, dynamically established. }
    \label{Fig1}
\end{figure}

\noindent \textbf{Introduction.---}
Many-body quantum systems with power-law decaying interactions appear in a variety of experimental setups \cite{RevModPhys.95.035002}, including  cavity-atom systems \cite{mivehvar2021}, trapped ions \cite{Britton2012,Feng2023} and Rydberg atoms \cite{Labuhn2016,Zeiher2017,Chen2023}. The interaction range, i.e., the power-law exponent $\alpha$ associated with the interaction potential [see sketch in Fig.~\ref{Fig1}(a)], can be controlled  \cite{Britton2012,Richerme2014}, and typically ranges from $\alpha=0$ for infinite-range interactions \cite{Britton2012}, to  $\alpha=6$ for van der Waals interactions \cite{saffman2010}, or can even be $\alpha\to\infty$ for nearest-neighbor forces  \cite{Jepsen2020}. The presence of long-range interactions allows for the emergence of novel physical effects  \cite{RevModPhys.95.035002,defenu2024outofequilibrium,Richerme2014,Defenu2021a,Zhang2017}, including nonlinear propagation of correlations or the inequivalence of thermodynamic ensembles. Harnessing long-range interactions is thus not only of fundamental interest but also of appeal for technological applications, e.g., to enhance correlations or accelerate entanglement spreading \cite{RevModPhys.95.035002,Jurcevic2014}. 

Long-range interacting systems are frequently considered in the context of spins, which are arranged on a $d$-dimensional lattice \cite{Nachtergaele2006,Storch_2015,LRB_long_Eisert}, and which interact pairwise with couplings following a power-law decay [cf.~Fig.~\ref{Fig1}(a)].
Exploring these systems is challenging (see also results in classical settings \cite{Cannas2000,Campa2009,Mori2010,Okuyama2023}), as state-of-the-art numerical methods face difficulties implementing long-range interactions and capturing correlations, in particular for open quantum systems and close to phase transitions  \cite{Czischek2018,Weimer2021,sulz2023numerical}. Moreover, only a few analytical results have been established beyond the case of {\it quadratic} theories \cite{Defenu2021a} or systems with infinite-range interactions \cite{RevModPhys_spohn,kastner2011,Paskauskas2012}, let alone for
nonequilibrium open quantum dynamics \cite{Alicki1983,Carollo2021,Boneberg2022,seetharam2022,carollo2023nongaussian,li2024,Merkli2012,Merkli2018,Fantechi2024, Mori2013}. 

In this paper, we rigorously derive the emergent dynamics of long-range interacting open quantum systems, in the {\it strong long-range regime} \cite{RevModPhys.95.035002}. We consider generic many-body systems, in arbitrary dimension, with power-law decaying interactions characterized by an exponent $\alpha\le d$ and a dynamical generator that is extensive in the particle number \cite{Kac_Factor,Defenu2021a,trigueros2023meanfield}. We demonstrate that, in the thermodynamic limit, the long-range dynamics is exactly captured by an effective nonlinear mean-field theory [cf.~Fig.~\ref{Fig1}(b)]. Our study goes beyond existing ones for closed quantum systems ~\cite{Granet2023} and classical ones~\cite{Cannas2000,Campa2009,Mori2010,Okuyama2023}.

Our rigorous results shed light on a number of intriguing phenomena: 
they explain the existence of {\it long-lived quasi-stationary states}  \cite{kastner2011,Paskauskas2012,Defenu2021a}, as fixed points of the nonlinear mean-field equations. They also predict the lack of {\it equilibration} in generic (unitary) long-range dynamics, as a consequence of the non-interacting character of the emergent dynamics. 
These findings are of immediate relevance for analyzing the behavior of a number of experimental platforms, including Rydberg atoms \cite{saffman2010,ding2024}, cavity-atom systems \cite{mivehvar2021,marsh2021}, and disordered long-range interacting systems that can host quantum associative memories \cite{rotondo2018,marsh2021,erba2021,fiorelli2023,marsh2024}.
\\

\noindent \textbf{Long-range open quantum systems.---}
We focus on many-body quantum systems defined on a $d$-dimensional regular lattice $\Gamma$, with $N=L^d$ sites [cf.~Fig.~\ref{Fig1}(a)]. 
The position of each site is identified by a vector $\textbf{r}=(\textbf{r}^1,\textbf{r}^2 ,\dots,\textbf{r}^d)^T$, whose entries can assume the values $\textbf{r}^\mu=1,2,\ldots,L$. 
The distance between two sites, $j$ and $k$, is given by the Euclidean distance $\|\textbf{r}_j-\textbf{r}_k\|_{\Gamma}$, where the subscript $\Gamma$ indicates that boundary conditions are accounted for. 
Each site is occupied by a $D$-level system and the Hilbert space associated with a subset of sites $\Lambda\subseteq \Gamma$ [cf.~Fig.~\ref{Fig1}(a)] is given by  $\mathcal{H}_\Lambda = \bigotimes_{j\in\Lambda} \mathbb{C}^D$.

The state of the many-body system is described by a density operator $\varrho_N$, such that $\varrho_N\ge0$ and ${\rm tr}_\Gamma\varrho_N=1$, where ${{\rm tr}_\Lambda}$ is the trace over the sites in $\Lambda$. Its evolution is implemented by the quantum master equation $\dot{\varrho}_{N,t}=\mathcal{L}_N[\varrho_{N,t}]$ \cite{lindblad1976,gorini1976,breuer2002theory}, with Lindblad generator $\mathcal{L}_N$ decomposing into a Hamiltonian and a dissipative term, as  $\mathcal{L}_N[\cdot]=-i[H_N,\cdot]+\mathcal{D}_N[\cdot]$. We take Hamiltonian operators of the form
\begin{equation}
H_N=\sum_{k\in \Gamma} h_k^{(k)}+\frac{1}{2c_\alpha^N }\sum_{j,k\in \Gamma,j\neq  k}\frac{V_{jk}^{(jk)}}{\|\textbf{r}_j-\textbf{r}_k\|_{\Gamma}^\alpha}\, . 
\label{Hamiltonian}
\end{equation}
The superscripts highlight the sites on which the operators act. The first sum contains (possibly inhomogeneous) single-site terms $h_k$, while the second one describes long-range interactions,  encoded in the (possibly inhomogeneous) two-body operators $V_{jk}$. The interaction strength decays algebraically with the distance, through the interaction-range exponent $\alpha$ [cf.~Fig.~\ref{Fig1}(a)]. For $\alpha\le d$, the double sum in Eq.~\eqref{Hamiltonian} becomes super-extensive in $N$, leading to unphysical behavior in the thermodynamic limit $N\to\infty$ \cite{kastner2011}. To prevent this, we include the so-called Kac factor $c_\alpha^N=\sum_{\textbf{r}\in[0,L-1]^d, \textbf{r} \neq 0} \|\textbf{r}\|_{\Gamma}^{-\alpha}$ [see Eq.~\eqref{Hamiltonian}], which restores the extensivity of the Hamiltonian \cite{Kac_Factor}. For large $N$, the Kac factor  scales  as \cite{kastner2011}
\begin{align}
    c_\alpha^N \overset{N\gg 1}{\propto} \left\{\begin{array}{cc}
        N^{1-\alpha/d} &  \alpha < d\, \\
         \log N& \alpha=d\, \\
         \mathrm{const.}& \alpha > d\, 
    \end{array}\right. \, .
    \label{Kac}
\end{align}
Dissipative dynamical effects are modeled via  the map
\begin{equation}
    \mathcal{D}_N\left[\cdot \right]=\sum_{\mu}\sum_{j,k\in\Gamma}\gamma_{jk}^\mu \left(L^{(j)}_\mu \cdot L_\mu^{(k)\, \dagger}-\frac{1}{2}\left\{L_\mu^{(k)\, \dagger}  L_\mu^{(j)} , \cdot\right\}\right)\,,
    \label{dissipator}
\end{equation}
which contains (possibly inhomogeneous) single-site and long-range terms. The index $\mu$ labels different processes represented by the operators $L_\mu^{(k)}$ and the matrices $\gamma_{jk}^\mu$ decompose as 
\begin{equation}
\gamma_{jk}^\mu  = 
     \kappa_j^\mu\delta_{jk}+
     \frac{w_{jk}^\mu}{ 2c_\alpha^N[\delta_{jk}+\|\textbf{r}_j-\textbf{r}_k\|^\alpha_\Gamma]} \, .
\label{Kossakowski}
\end{equation}
The rates $\kappa_j^\mu$ are local dissipation rates, while the matrices $w^\mu_{jk}$ specify long-range collective dissipative effects. The generator $\mathcal{L}_N$ must implement a completely positive dynamics, which is for instance ensured by  $\gamma^\mu\ge0$, $\forall \mu$. 
It can also be divided into local and nonlocal terms as [see details in Ref.~\cite{sm} and Fig.~\ref{Fig1}(a)]
\begin{equation}\label{eq:gen_decomposed}
  \mathcal{L}_N = \sum_{k\in\Gamma}\mathcal{L}_k^{(k)} + \frac{1}{2c_\alpha^N} \sum_{j,k\in\Gamma}\frac{\mathcal{L}_{jk}^{(jk)}}{[\delta_{jk}+\|\textbf{r}_j-\textbf{r}_k\|^\alpha_\Gamma]}\,  .
\end{equation}
The second term features an unconstrained double sum, so that the factor $1/2$ avoids  double counting. We assume $h_k$, $V_{jk}$, $\gamma^\mu$ and $L_\mu^{(k)}$ to be bounded.\\

\noindent \textbf{Local mean-field dynamics.---} Accessible information about many-body systems is contained in reduced quantum states, describing finite subsets of particles [cf.~Fig.~\ref{Fig1}(a)]. For instance, sample-average properties, e.g., order parameters, can be computed through single-particle states while two-body correlations through two-particle states. In what follows, we focus on the reduced state of a generic finite subset of particles  $\Lambda$ and demonstrate that, in the regime $\alpha\le d$ and in the thermodynamic limit, it obeys a nonlinear mean-field dynamics  [cf.~Fig.~\ref{Fig1}(b)]. To show this fact, we give here a heuristic argument (see also Ref.~\cite{benedikter2016}) while we provide a rigorous proof in Ref.~\cite{sm}. 

Let us consider a finite subset of sites, $\Lambda\subset\Gamma$, and denote the associated reduced state as  $\varrho_{N,t}^{(\Lambda)}={\rm tr}_{\Lambda'}\varrho_{N,t}$, where $\Lambda'$ indicates the complement of $\Lambda$. By taking the time derivative  and recalling Eq.~\eqref{eq:gen_decomposed}, we find 
\begin{equation}
\begin{split}
\dot{\varrho}_{N,t}^{(\Lambda)}&=\sum_{k\in\Lambda}\mathcal{L}_{k}^{(k)}[\varrho_{N,t}^{(\Lambda)}]+\frac{1}{c_\alpha^N}\sum_{j\in \Lambda, k\in \Lambda'} \mathcal{C}_{jk}^N[{\varrho}_{N,t}^{(\Lambda\cup \{k\})}]
\\
&+\frac{1}{2c_\alpha^N} \sum_{j,k\in \Lambda} \frac{\mathcal{L}_{jk}^{(jk)}[{\varrho}_{N,t}^{(\Lambda)}]}{[\delta_{jk}+\|\textbf{r}_j-\textbf{r}_k\|^\alpha_\Gamma]}\,  .
\end{split}
\label{der_Lambda}
\end{equation}
The first term on the right-hand side of the equation includes local dynamical contributions, while the third one contains interactions among particles in $\Lambda$. The second term accounts for interactions between the particles in $\Lambda$ and those in the remainder, $\Gamma \setminus \Lambda$, of the many-body system. Such an interaction is captured by the ``collision operator" \cite{RevModPhys_spohn} 
\begin{equation}
\mathcal{C}_{jk}^N[\varrho_N^{(\Lambda\cup \{k\})}]=\frac{1}{\|\textbf{r}_j-\textbf{r}_k\|^\alpha_\Gamma}{\rm tr}_k\left(\mathcal{L}_{jk}^{(jk)}[{\varrho}_{N,t}^{(\Lambda\cup \{k\})}]\right)\, , 
\label{collision}
\end{equation}
for $j\in \Lambda$, $k\in \Lambda'$. The relation in Eq.~\eqref{der_Lambda} defines a hierarchy of differential equations, manifesting in  ${\varrho}_{N,t}^{(\Lambda)}$ depending on the reduced states ${\varrho}_{N,t}^{(\Lambda\cup \{k\})}$ involving the particles in $\Lambda$ plus a particle in $\Lambda'$ \cite{RevModPhys_spohn,Alicki1983,Laszlo2007,benedikter2016,Paskauskas2012}.

As shown by Eq.~\eqref{Kac}, in the strong long-range regime $\alpha\le d$ \cite{RevModPhys.95.035002}, the factor $c_\alpha^N$ diverges. This suggests that, the term in the second line of Eq.~\eqref{der_Lambda}, solely involving finite sums, is vanishing for large $N$. On the other hand, the collision term survives as it is averaged over an extensive number of particles.  For $N\gg1$, we thus  have 
\begin{equation}
\dot{\varrho}_{N,t}^{(\Lambda)}\approx \sum_{k\in\Lambda}\mathcal{L}_{k}^{(k)}[\varrho_{N,t}^{(\Lambda)}]+\sum_{j\in\Lambda,k\in \Lambda'}\frac{1}{c_\alpha^N}\, \mathcal{C}_{jk}^N[{\varrho}_{N,t}^{(\Lambda\cup \{k\})}]\, . 
\label{inf_eq}
\end{equation}
Since the equation is valid for finite $\Lambda$ and large $N$, we can additionally extend the sum over $k$, from $k\notin \Lambda$ to $k\neq j$. This introduces a new hierarchy of equations, which is solved by reduced states  ${\varrho}_{N,t}^{(\Lambda)}\approx \bigotimes_{j\in \Lambda}\tilde{\varrho}_{N,t}^{(j)}$, with the single-particle states $\tilde{\varrho}_{N,t}^{(j)}$ obeying suitable dynamical equations [see Eq.~\eqref{eq:effective_generator} below].

\begin{theorem}\label{theorem}
Let us consider an initial state $\varrho_N=\bigotimes_{j\in\Gamma}\varrho_N^{(j)}$ and the generator in Eq.~\eqref{eq:gen_decomposed}. For $\alpha\le d$, given a finite set $\Lambda$, we have
\begin{equation}
\underset{N\to\infty}{\lim} \Big[\varrho_{N,t}^{(\Lambda)} - \underset{j\in\Lambda}{\bigotimes} \tilde{\varrho}_{N,t}^{(j)}\Big]=0\, ,
\label{th1}
\end{equation}
in trace-norm. Here,  $\tilde{\varrho}_{N,t}^{(j)}$ obeys the equation
\begin{equation}\label{eq:effective_generator}
  \dot{\tilde{\varrho}}_{N,t}^{(j)}  =  \mathcal{L}_j[{\tilde{\varrho}}_{N,t}^{(j)}  ]+\sum_{k\in\Gamma,k\neq j }\frac{{\rm tr}_2(\mathcal{L}_{jk}[\tilde{\varrho}_{N,t}^{(j)}\otimes \tilde{\varrho}_{N,t}^{(k)}])}{c_\alpha^N \|{\bf r}_j-{\bf r}_k\|^\alpha_\Gamma}\, ,
\end{equation}
with  $\tilde{\varrho}_{N,t=0}^{(j)}=\varrho_N^{(j)}$. The map $\mathcal{L}_j$ acts on  single-particle reduced states, $\mathcal{L}_{jk}$ acts on the tensor product of two single-particle states, and ${\rm tr}_2$ is the trace over the second entry of the tensor product. 
\end{theorem} 
\textit{Proof.} The proof relies on constructing perturbation-series expansions, in terms of the collision operators, for the solutions of Eq.~\eqref{der_Lambda}, Eq.~\eqref{inf_eq}, and  Eq.~\eqref{eq:effective_generator}. It then shows that these series converge (uniformly in $N$ and in trace norm) \cite{RevModPhys_spohn} and that they are equivalent in the thermodynamic limit (see details in Ref.~\cite{sm}). \qed 

Our theorem shows that, for large $N$, reduced states of the many-body quantum system remain in product form, if they initially are, and are fully determined by the single-particle states  $\tilde{\varrho}_{N,t}^{(j)}$. The latter evolve through a dissipative Hartree-like \cite{RevModPhys_spohn,Merkli2012,benedikter2016} dynamics [cf.~Eq.~\eqref{eq:effective_generator}]. 
Contrary to a standard  (permutation-invariant) mean-field scenario, the generic situation  considered so far does not allow for an immediate statement on the convergence, with $N$, of $\tilde{\varrho}_{N,t}^{(j)}$ to a well-defined limit $\tilde{\varrho}_{\infty,t}^{(j)}$. This is due to the allowed inhomogeneous parameters, or disorder, in the dynamical generator [cf.~Eq.~\eqref{Hamiltonian} and Eq.~\eqref{Kossakowski}]. Moreover, it is intrinsically related to the fact that long-range interactions are not permutation invariant. As a consequence, the convergence of reduced single-particle states is strictly tied to the convergence of the average over collisions in Eq.~\eqref{eq:effective_generator} (see Proposition \ref{proposition} below), which is not obvious as in permutation-invariant cases. Even in the most general case, our theorem however permits the efficient analysis of the  system as well as of convergence of the reduced single-particle states, e.g., via numerical integration of Eq.~\eqref{eq:effective_generator} for increasing $N$.

To make analytical progress, we  focus instead on homogeneous single-site and interaction terms. That is, we consider $\mathcal{L}_k=\mathcal{L}_1$, $\forall k$, and $\mathcal{L}_{jk}=\mathcal{L}_{12}$,  $\forall j\neq k$. In this setting, we prove convergence of the reduced states with $N$ and discuss the role of the boundary conditions. \\

\noindent \textbf{Impact of boundary conditions.---} For short-range interacting systems, the lattice boundary conditions are  irrelevant, in the thermodynamic limit, when considering bulk properties. As we shall see, instead, in the presence of strong long-range interactions, boundary conditions have a dramatic impact.

The case of periodic boundary conditions is straightforward to treat with our theorem and becomes equivalent to typical mean-field scenarios. In fact, assuming a  translation-invariant initial product state, the periodic boundary conditions ensure a translation-invariant system. Our theorem shows that reduced states remain in product form, so that the translation invariance of the system is 
``promoted", in the thermodynamic limit, to an emergent permutation invariance. Rigorously, this fact can be established by considering that a translation-invariant ansatz for reduced states, i.e.,  $\tilde{\varrho}_{N,t}^{(j)} =\tilde{\varrho}_{N,t}^{(k)}$, $\forall j,k$, solves Eq.~\eqref{eq:effective_generator} in the case of periodic boundary conditions. We thus have that  $\tilde{\varrho}_{N,t}^{(j)}\to \tilde{\varrho}_{\infty,t}$, which does not depend on $j$ and obeys the nonlinear differential equation
\begin{equation}
\dot{\tilde{\varrho}}_{\infty,t}=\mathcal{L}_1[\tilde{\varrho}_{\infty,t}]+{\rm tr}_2\left(\mathcal{L}_{12}[\tilde{\varrho}_{\infty,t}\otimes \tilde{\varrho}_{\infty,t}]\right)\, , 
\label{Hart-like}
\end{equation}
reminiscent of infinite-range interacting systems  \cite{Alicki1983}. 

To address the case of open boundary conditions, we introduce a rescaled spatial coordinate $\textbf{x}_j=\textbf{r}_j/L$ and  relabel the reduced  states as $\tilde{\varrho}_{N,t}^{(j)}\leftrightarrow \tilde{\varrho}_{N,t}^{(\textbf{x}_j)}$. For homogeneous dynamical generators, the continuity of reduced single-particle states in the rescaled spatial coordinate is preserved in time. This allows us to recast the average over collisions in Eq.~\eqref{eq:effective_generator} as a Riemann sum. In the thermodynamic limit $N\to\infty$, the latter converges to an  integral, as formalized in the following proposition (see proof in Ref.~\cite{sm}).

\begin{figure}[t!]
    \centering
    \includegraphics[width=\columnwidth]{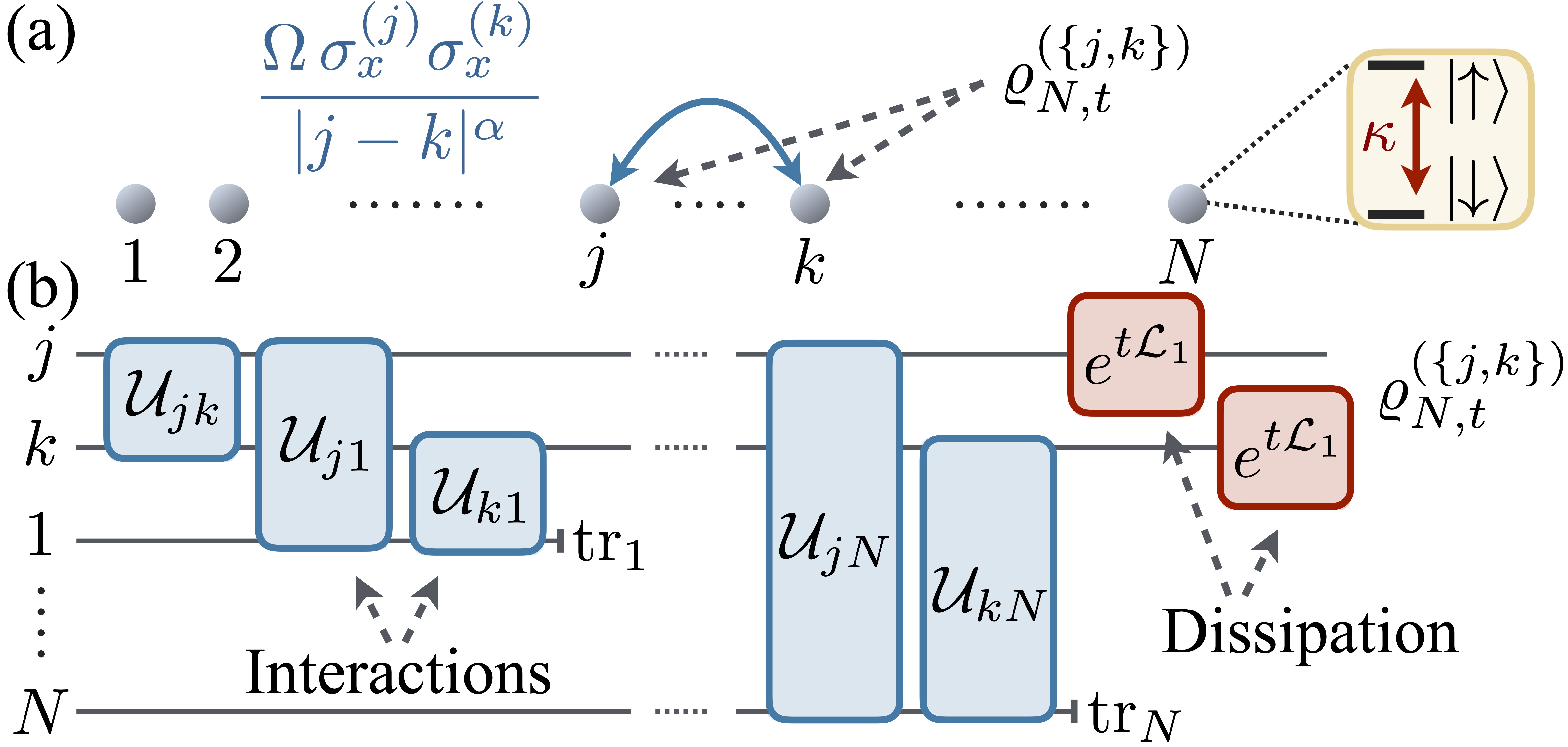}
    \caption{{\bf Long-range dissipative quench.} (a) One-dimensional open chain composed of $N$ spin-$1/2$ particles, featuring power-law   interactions with coupling strength $\Omega$. Local dissipation implements transitions between spin states, $\ket{\downarrow}\leftrightarrow\ket{\uparrow}$, at rate $\kappa$. We focus on the evolution of two sites. (b) Collision-model circuit used for numerical simulations. The reduced two-spin state $\varrho_{N,t}^{(\{j,k\})}$ is obtained by applying a sequence of unitary gates and  two additional dissipative gates on the spins of interest. After each interaction of the two spins with an external one, the latter can be traced out.  }
    \label{Fig2}
\end{figure}

\begin{proposition}
\label{proposition}
  Let us consider the generator in Eq.~\eqref{eq:gen_decomposed}, with $\alpha<d$ and homogeneous single-site and interaction terms. Let us also assume an initial product state $\varrho_N=\bigotimes_{j\in\Gamma}\varrho_N^{({\bf x}_j)}$, which is continuous in the rescaled coordinate ${\bf x}_j$ in the thermodynamic limit. The reduced state $\tilde{\varrho}_{N,t}^{({\bf x}_j)}$ converges, in the thermodynamic limit, to  $\tilde{\varrho}_{\infty,t}^{({\bf x})}$, with $\lim_{N\to\infty} {\bf x}_j={\bf x}$,  satisfying
\begin{equation}\label{eq:dyn_proposition}
 \dot{\tilde{\varrho}}_{\infty,t}^{({\bf x})} = \mathcal{L}_1[  \tilde{\varrho}_{\infty,t}^{({\bf x})}] +\theta(\alpha) \!\int_0^1\! {\rm d}{\bf y} \frac{\mathrm{tr}_{2}  (\mathcal{L}_{12}[  \tilde{\varrho}_{\infty,t}^{({\bf x})} \otimes \tilde{\varrho}_{\infty,t}^{({\bf y})}]) }{\|{\bf x}-{\bf y}\|_\Gamma^\alpha}\, ,
\end{equation}
where $\theta(\alpha) = \lim_{N\to \infty} {N^{1-\alpha/d}}/{c_\alpha^N}$.
\end{proposition}
The presence of long-range interactions is such that, the dynamics of reduced single-particle states is ``instantaneously" affected by the state of all other sites in the lattice [accounted for by the integral in Eq.~\eqref{eq:dyn_proposition}], even of those infinitely far apart. Note that, the case $\alpha=d$ is not included since in this case  the average over collisions is not equivalent to a Riemann sum [cf.~Eq.~\eqref{Kac}].\\

\noindent \textbf{Long-range dissipative quench.---}
To benchmark our derivation, we focus on a dissipative instance of the paradigmatic long-range  Ising model, see, e.g., Refs.~\cite{foss-feig2013,Richerme2014,Defenu2021a}. Considering such a system further allows us to demonstrate how our results can shed new insights on the dynamical behavior of long-range interacting systems, e.g., on the emergence of long-lived quasi-stationary states \cite{kastner2011,Paskauskas2012,Defenu2021a}. 

We take a spin-$1/2$ chain ($d=1$) with open boundaries, as illustrated in Fig.~\ref{Fig2}(a). The unitary part of dynamics is governed by a  Hamiltonian like the one in Eq.~\eqref{Hamiltonian},  with  $h_k=0$ and fixing $V_{12}^{(jk)}=\Omega\sigma^{(j)}_x\sigma^{(k)}_x$. The dissipative contribution is local, with 
\begin{equation}\label{eq:dissipator_quench}
    \mathcal{L}_1^{(j)}[\cdot]= \kappa\left(\sigma_x^{(j)} \cdot \sigma_x^{(j)}- \frac{1}{2}\left\{\sigma_x^{(j)}\sigma_x^{(j)}, \cdot\right\}\right)\, ,
\end{equation}
and accounts for incoherent transitions between the two spin states [cf.~Fig.~\ref{Fig2}(a)]. For such model, the dissipative contributions and the Hamiltonian interaction terms commute among themselves. This fact allows us to simulate the evolution of reduced states, by means of a collision-model circuit \cite{CICCARELLO20221,Cattaneo2022} like the one sketched in Fig.~\ref{Fig2}(b), for up to $10^7$ spin particles. Such  large numbers of particles are  needed to observe the onset of the mean-field dynamics throughout the whole regime $\alpha\in[0,d]$. 

\begin{figure}[t!]
    \centering
     \includegraphics[width=\columnwidth]{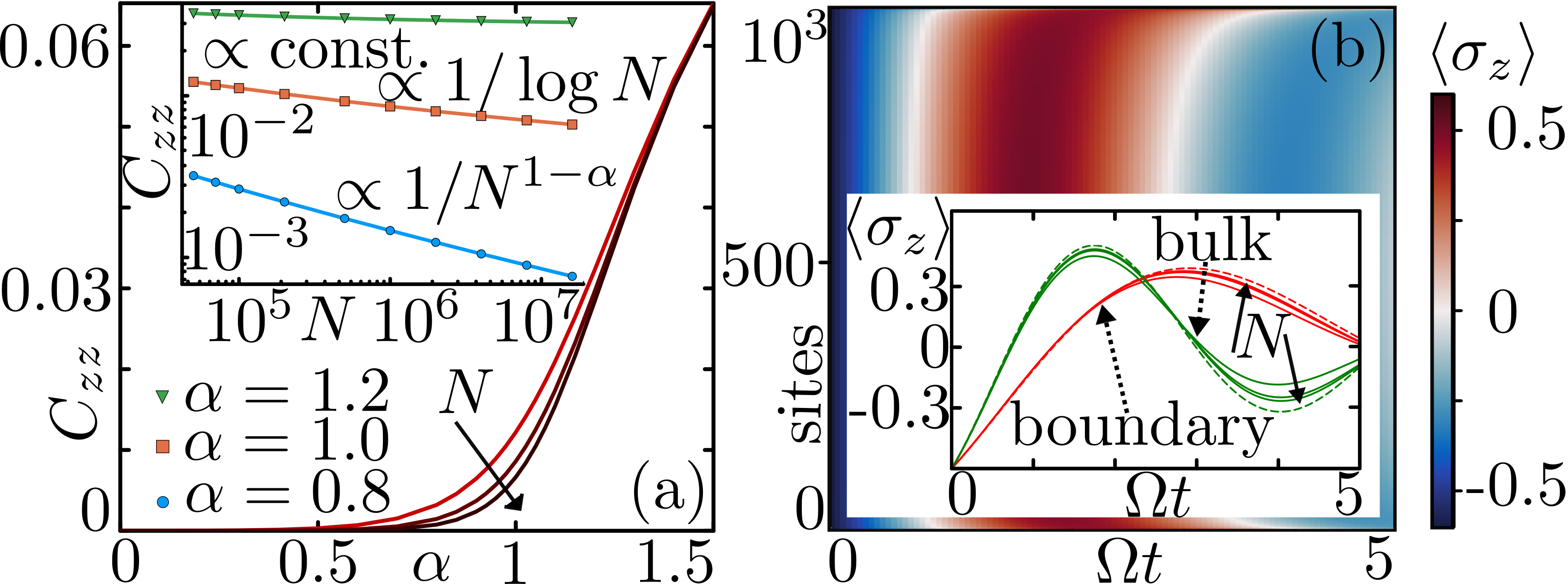}
    \caption{{\bf Numerical benchmark.} (a) Correlation $C_{zz}$ (see main text) of spins $j$ and $j+1$, with $j=N/2$. The plot shows different system sizes,  $N=5\times10^{4}, 1\times10^{6}, 16\times10^{6}$, for $\kappa=0.1\Omega$ and $\Omega t=0.5$. The inset shows $C_{zz}$ for up to $N=16\times10^6$ and $\alpha=0.8,1,1.2$. (b) Spin magnetization $\langle \sigma_z\rangle$ as a function of time and lattice site, for $N=1000$, $\kappa=0.1\Omega$ and $\alpha=0.8$. The inset shows  $\langle \sigma_z \rangle$ for a particle in the bulk ($j=N/2$) and one at the boundary ($j=1$). The dashed line is the effective dynamics   [cf.~Eq.~\eqref{eq:eff_dyn_quench}] while  solid lines are numerical results for $N=1000,5000,10000$. For both panels, the initial state is $\varrho_N=\otimes_{k=1}^N\varrho_1$ with $\varrho_1=[1,1-i;1+i,4]/5$. }
    \label{Fig3}
\end{figure}

First, we show that, for increasing $N$, two-spin reduced states converge to a product state. To this end, we consider the correlation $C_{zz}= \langle \sigma_z^{(j)} \sigma_z^{(j+1)} \rangle - \langle \sigma_z^{(j)} \rangle \langle  \sigma_z^{(j+1)} \rangle$, between adjacent particles at the center of the chain ($j=N/2$) and at a fixed time. This quantity, which effectively acts as an ``order parameter" for mean-field behavior, is shown in Fig.~\ref{Fig3}(a) for different values of $\alpha$.  For $\alpha\ll1$, the correlation is essentially zero for the considered system sizes. In the regime $\alpha\in[0.5, 1]$,  we instead observe a nonzero correlation. The latter is, however, only due to finite-size effects as demonstrated by the scaling shown in the inset. For $\alpha>1$, the correlation instead converges to a finite value in the thermodynamic limit [see inset of Fig.~\ref{Fig3}(a)]. In Ref.~\cite{sm}, we show additional results on the quantum mutual information.

We then perform a direct computation of the inhomogeneous mean-field equation in Eq.~\eqref{eq:dyn_proposition}.  Considering our theorem [cf.~Eq.~\eqref{eq:effective_generator}], for the model at hand the collision operator   reduces to a nonlinear Hamiltonian contribution $\propto {\rm tr}(\sigma_x\tilde{\varrho}_{N,t}^{(\textbf{x}_k)}) [\sigma_x,\tilde{\varrho}_{N,t}^{(\textbf{x}_j)}]$. Since $\sigma_x$ is conserved, the trace is equal to the initial expectation value $\langle \sigma_x\rangle$. The average over collisions in Eq.~\eqref{eq:effective_generator} can thus be written as a Riemann sum solely involving the power-law function. As such, we analytically calculate   
\begin{align}\label{eq:eff_dyn_quench}
    \dot{\varrho}^{(\textbf{x})}_{\infty,t} = -i \phi(\textbf{x}) [\sigma_x,{\varrho}^{(\textbf{x})}_{\infty,t} ] + \mathcal{L}_1[{\varrho}^{(\textbf{x})}_{\infty,t}]\, ,
\end{align}
with $\phi(\textbf{x})=\langle \sigma_x\rangle \Omega\theta(\alpha)\int_0^1 {\rm d}\textbf{y}\, |\textbf{x}-\textbf{y}|^{-\alpha}$. Such an equation is  benchmarked in Fig.~\ref{Fig3}(b). 

The nonlinear and non-interacting character of the dynamics governed by Eq.~\eqref{eq:eff_dyn_quench}  provides simple, yet rigorous, insights on the emergence of the long-lived quasi-stationary states observed in Refs.~\cite{kastner2011,Defenu2021a}. There, a unitary quench from a state with $\langle \sigma_x\rangle =0$ was considered. Our results imply that, in the thermodynamic limit, the system must obey  Eq.~\eqref{eq:eff_dyn_quench}, with $\kappa=0$, so that an initial state with $\langle \sigma_x\rangle=0$  is a fixed point of the emergent dynamics. Therefore, for finite systems, the quantum many-body state must necessarily become longer and longer lived upon increasing the number of particles. [Fig.~\ref{Fig3}(b) displays  dynamics since we consider a state with $\langle \sigma_x\rangle \neq 0$.] This example also illustrates the non-commuting character of the limits $N\to\infty$ and $t\to\infty$.\\ 

\noindent \textbf{Discussion.---}
While exact results exist for Hamiltonian systems with collective~\cite{RevModPhys_spohn,Granet2023} and power-law interactions~\cite{Mori2019,kastner2011,Paskauskas2012,Defenu2021a,Granet2023} as well as for open quantum systems with collective interactions~\cite{Alicki1983,Carollo2021,seetharam2022,carollo2023nongaussian,li2024,Merkli2012,Merkli2018,Mori2013}, our theorem goes beyond these  cases. Furthermore, our work differs significantly from the ones mentioned above since Theorem~\ref{theorem}, which takes inspiration from Refs.~\cite{RevModPhys_spohn,Alicki1983,benedikter2016}, is not restricted to homogeneous interactions, but covers the cases of disordered single-site and interacting terms as well as of generic inhomogeneous dynamics. Importantly, it allows for the exploration of strong long-range interacting dynamics with a linear complexity in the system size  [cf.~Eq.~\eqref{eq:effective_generator}]. Within this formalism, it is possible to describe nonequilibrium phase transitions and spontaneous symmetry breaking (see discussion in Ref.~\cite{sm}). Our proposition demonstrates the impact of boundary conditions on the emergent dynamics. For periodic boundaries and translation-invariant dynamical generators, the system becomes permutation invariant in the thermodynamic limit. For open boundaries, it is instead described by an inhomogeneous mean-field theory (see, e.g.,  Refs.~\cite{fannes1982,Duffield1992,Tschopp2020,puschmann2021} for inhomogeneous mean-field equations in related contexts), where contrary to recent results~\cite{Iemini2024} the inhomogeneity emerges from the dynamics itself and is not a remnant of the chosen initial state. While  Eqs.~(\ref{eq:effective_generator}-\ref{eq:dyn_proposition}) may look abstract at first, they simply entail  that the relevant equations of motion can be found by considering a product-state ansatz. We stress that our results are valid for reduced states and do not preclude the possibility of highly entangled full many-body states \cite{Koffel2012,Cadarso2013} nor of measurement-induced phase transitions  \cite{Passarelli2024}. We also note that they can be exploited to compute the expectation value of quasi-local operators and not only of strictly local ones \cite{Bratelli1981, sm}.

The collision term, i.e., the second term on the right hand side of Eq.~\eqref{eq:effective_generator}, always gives rise to a nonlinear single-site Hamiltonian term [see, e.g., Eq.~\eqref{eq:eff_dyn_quench}], even when it stems from long-range dissipative contributions  \cite{Alicki1983,benatti2018}. In the case of unitary many-body systems with strong long-range interactions (i.e., $\gamma^\mu_{jk}\equiv0$), our results thus generically predict the absence of equilibration of local observables to a stationary value, given that the emergent dynamics is governed by an effective non-interacting Hamiltonian. The considered example allowed us to  understand the general mechanism behind the emergence of long-lived quasi-stationary states in long-range systems \cite{kastner2011,Paskauskas2012,Defenu2021a}. This  phenomenon occurs when the finite-$N$ state is associated, in the thermodynamic limit, with a fixed point of the emergent mean-field equations.\\

The code and the data for the dissipative quench described in the manuscript are available on Zenodo \cite{mattes_2024_zenodo}.\\

\noindent \textbf{Acknowledgments.---} 
We are grateful to Alvise Bastianello and Stefan Teufel for useful comments. 
We acknowledge funding from the Deutsche Forschungsgemeinschaft (DFG, German Research Foundation) under Project No.~435696605 and through the Research Unit FOR 5413/1, Grant No.~465199066, through the Research Unit FOR 5522/1, Grant No.~499180199. This project has also received funding from the European Union’s Horizon Europe research and innovation program under Grant Agreement No.~101046968 (BRISQ). This work was funded within the QuantERA II Programme (project CoQuaDis, DFG Grant No. 532763411) that has received funding from the EU H2020 research and innovation programme under GA No. 101017733. FC~is indebted to the Baden-W\"urttemberg Stiftung for the financial support of this research project by the Eliteprogramme for Postdocs.
\bibliography{refs.bib}
\newpage
\setcounter{equation}{0}
\setcounter{figure}{0}
\setcounter{table}{0}
\makeatletter
\renewcommand{\theequation}{S\arabic{equation}}
\renewcommand{\thefigure}{S\arabic{figure}}
\makeatletter

\onecolumngrid
\newpage

\setcounter{page}{1}
\begin{center}
{\Large SUPPLEMENTAL  MATERIAL}
\end{center}
\begin{center}
\vspace{0.8cm}
{\Large Long-range interacting systems are locally non-interacting}
\end{center}
\begin{center}
Robert Mattes,$^{1}$ Igor Lesanovsky,$^{1,2}$ and Federico Carollo$^{3}$
\end{center}
\begin{center}
$^1${\em Institut f\"ur Theoretische Physik, Universit\"at T\"ubingen,}\\
{\em Auf der Morgenstelle 14, 72076 T\"ubingen, Germany}\\
$^2${\em School of Physics and Astronomy and Centre for the Mathematics}\\
{\em and Theoretical Physics of Quantum Non-Equilibrium Systems,}\\
{\em  The University of Nottingham, Nottingham, NG7 2RD, United Kingdom}\\
$^3${\em Centre for Fluid and Complex Systems, Coventry University, Coventry, CV1 2TT, United Kingdom}

\end{center}
\section{I. Proof of main theorem}
In this section, we provide the detailed proof of our Theorem \ref{theorem}. In preparation to that, we give an explicit definition of the relevant maps forming the dynamical generators that are exploited in the derivation. We further compute the dynamical equation for generic reduced quantum states.

As reported in Eq.~\eqref{eq:gen_decomposed} of the main text, the full dynamical generator $\mathcal{L}_N$ can be decomposed into a part containing single-site terms 
\begin{equation}
    \mathcal{L}^{(k)}_k[\cdot]=-i \left[h_k^{(k)},\cdot\right]+\sum_\mu  \kappa_k^\mu  \left(L^{(k)}_\mu \cdot L_\mu^{(k)\, \dagger}-\frac{1}{2}\left\{L_\mu^{(k)\, \dagger}  L_\mu^{(k)} , \cdot\right\}\right)\, ,
    \label{L1}
\end{equation}
and one collecting long-range interactions   
\begin{equation}
\begin{split}
    \mathcal{L}^{(jk)}_{jk}[\cdot]&=-\frac{i(1-\delta_{jk})}{2} \left[V_{jk}^{(jk)}{+ V_{kj}^{(jk)}},\cdot\right]+\\
    &+\sum_\mu\left[  \frac{w_{jk}^\mu}{2}  \left(L^{(j)}_\mu \cdot L_\mu^{(k) \dagger}-\frac{1}{2}\left\{L_\mu^{(k) \dagger}  L_\mu^{(j)} , \cdot\right\}\right)+ \frac{w_{kj}^\mu}{2}  \left(L^{(k)}_\mu \cdot L_\mu^{(j) \dagger}-\frac{1}{2}\left\{L_\mu^{(j) \dagger}  L_\mu^{(k)} , \cdot\right\}\right)\right] .
    \end{split}
    \label{L_jk}
\end{equation}
Note that, $\mathcal{L}_{jk}=\mathcal{L}_{kj}$ and that $\mathcal{L}_{kk}^{(kk)}$ is also a map acting nontrivially only on site $k$, but it is different from the map $\mathcal{L}^{(k)}_k$ as it considers terms coming from the long-range interactions. To simplify the notation, for the dynamical generators we use the same symbol for both when they act on the full system and on a part of it. The actual support of the generators in the different situations will be evident from the support of the operator they act upon. For instance, in the expression $\mathcal{L}_{jk}^{(jk)}[\varrho_N]$ the generator $\mathcal{L}_{jk}^{(jk)}$ acts on the whole system (but nontrivially only on site $j$ and $k$), while in the expression $\mathcal{L}_{jk}^{(jk)}[\varrho_N^{(\Lambda)}]$ it  solely acts on the sites defining $\Lambda$.

With the above  maps, we can define the ``reduced" dynamical generator 
\begin{equation}\label{eq:gen_lambda}
    \mathcal{L}^{(\Lambda)}=\sum_{k\in \Lambda} \mathcal{L}^{(k)}_k +\frac{1}{2c_\alpha^N} \sum_{j,k\in \Lambda} \frac{\mathcal{L}_{jk}^{(jk)}}{[\delta_{jk}+\|\textbf{r}_j-\textbf{r}_k\|^\alpha_\Gamma]} \, ,
\end{equation}
acting nontrivially only on sites in  $\Lambda$. For completeness, we note that the Euclidean distance between two lattice sites $j$ and $k$ is given by
$
    \|\textbf{r}_j-\textbf{r}_k\|_\Gamma = \sqrt{\sum_{\mu=1}^d [\mathrm{dist}(\textbf{r}_j^\mu,\textbf{r}_k^\mu)]^2}\, ,
$
where $\mathrm{dist}(\textbf{r}_j^\mu,\textbf{r}_k^\mu)=|\textbf{r}_j^\mu-\textbf{r}_k^\mu|$, for open boundary conditions, while it is $\mathrm{dist}(\textbf{r}_j^\mu,\textbf{r}_k^\mu)=\mathrm{min}(|\textbf{r}_j^\mu-\textbf{r}_k^\mu|,L-|\textbf{r}_j^\mu-\textbf{r}_k^\mu|)$, for periodic boundary ones.
\\

The central objects of our theorem are  the reduced quantum states $\varrho_{N,t}^{(\Lambda)}={\rm tr}_{\Lambda'}\left(\varrho_{N,t}\right)$, with $\varrho_{N,t}=e^{t\mathcal{L}_N}[\varrho_{N}]$ being the time-evolved state of the full many-body system. These states contain the full information about generic sets of sites $\Lambda$. Taking the  time derivative of  $\varrho_{N,t}^{(\Lambda)}$ we find
\begin{equation}
\dot{\varrho}_{N,t}^{(\Lambda)}=\mathcal{L}^{(\Lambda)}[{\varrho}_{N,t}^{(\Lambda)}]+\frac{1}{c_\alpha^N}\sum_{j\in \Lambda}\sum_{k\notin \Lambda}\frac{{\rm tr}_{k}\left(\mathcal{L}_{jk}^{(jk)}[{\varrho}_{N,t}^{(\Lambda\cup\{k\})}]\right)}{\|{\bf r}_j-{\bf r}_k\|_\Gamma^\alpha}\, ,
    \label{der-red-state}
\end{equation}
which follows from considering the above definitions and that ${\rm tr}_{\Lambda'}\left(\mathcal{L}_{jk}^{(jk)}[\varrho_{N,t}]\right)=0$, if both $j,k\in \Lambda'$, due to the fact that $\mathcal{L}_{jk}^{(jk)}$ are trace-preserving. In the equation above, ${\varrho}_{N,t}^{(\Lambda\cup\{k\})}$ is the reduced state for the sites in $\Lambda$ plus  the site $k\notin \Lambda$. With these preliminary considerations, we are now ready to prove our theorem.\\ 

\noindent {\it Proof.}
The formal solution of Eq.~\eqref{der-red-state} can be written as 
\begin{equation}\label{eq:formal_sol}
    \varrho_{N,t}^{(\Lambda)}=e^{t\mathcal{L}^{(\Lambda)}}[\varrho_{N,0}^{(\Lambda)}]+\frac{1}{c_\alpha^N}\int_0^{t}{\rm d}t_1 \sum_{j\in\Lambda}\sum_{k\notin \Lambda} \frac{ e^{(t-t_1)\mathcal{L}^{(\Lambda)}} \circ{\rm tr}_k\circ\mathcal{L}_{jk}^{(jk)}\left[\varrho_{N,t_1}^{(\Lambda\cup \{k\})}\right]}{\|{\bf r}_j-{\bf r}_k\|_\Gamma^\alpha}\, ,
\end{equation}
where we use the symbol $\circ$ to denote composition of maps (the partial trace is indeed also a map acting on operators). 
The above relation can be iterated, by substituting it into itself, to define a perturbation-series expansion for $\varrho_{N,t}^{(\Lambda)}$, in terms of a sequence of collisions. The perturbation series reads
\begin{equation}
\varrho_{N,t}^{(\Lambda)}=\sum_{m=0}^{N-|\Lambda|} \int_0^{t}{\rm d}t_1\int_0^{t_1}{\rm d}t_2\dots \int_0^{t_{m-1}}{\rm d}t_m \, T_{m}\, ,
\label{series_T}
\end{equation}
where we have 
\begin{equation}
\begin{split}
T_m&=\frac{1}{(c_\alpha^N)^m}\sum_{j_1\in\Lambda}\sum_{k_1\notin\Lambda}\sum_{j_2\in \Lambda\cup \{k_1\}}\sum_{k_2\notin \Lambda\cup \{k_1\}}\dots \sum_{j_m\in \Lambda\cup \{k_1,k_2,\dots k_{m-1}\}}\sum_{k_m\notin \Lambda\cup \{k_1,k_2,\dots k_{m-1}\}} \Phi_m \left[\varrho_{N,0}^{(\Lambda\cup \{k_1,\dots k_m\})}\right]
\end{split}
    \label{series_term}
\end{equation}
and  
\begin{equation}
\begin{split}
\Phi_m \left[\varrho_{N,0}^{(\Lambda\cup \{k_1,\dots k_m\})}\right] =\frac{1}{\|{\bf r}_{j_1}-{\bf r}_{k_1}\|^\alpha_\Gamma\|{\bf r}_{j_2}-{\bf r}_{k_2}\|^\alpha_\Gamma\dots \|{\bf r}_{j_m}-{\bf r}_{k_m}\|^\alpha_\Gamma}e^{(t-t_1)\mathcal{L}^{(\Lambda)}}\circ{\rm tr}_{k_1}\circ\mathcal{L}_{j_1k_1}^{(j_1k_1)}\circ e^{(t_1-t_2)\mathcal{L}^{(\Lambda\cup \{k_1\})}}\circ \dots \\
\dots \circ {\rm tr}_{k_m}\circ \mathcal{L}_{j_mk_m}^{(j_mk_m)} \circ e^{t_m\mathcal{L}^{(\Lambda\cup \{k_1,\dots k_m\})}}\left[\varrho_{N,0}^{(\Lambda\cup \{k_1,\dots k_m\})}\right]\, . 
\end{split}
    \label{series_term2}
\end{equation}
As shown in Lemma \ref{lemma1}, each term $T_m$ is bounded by $\|T_m\|_{\rm tr}\le (2^d C)^m n(n+1)\dots (n+m-1)$ and the whole series converges, in trace norm, for $t<t_0 = 2^d C$. Note that, since the proof of the Lemma solely exploits that $\varrho_{N,0}^{(\Lambda\cup \{k_1,\dots k_m\})}$ is trace class and since this holds true also for $\varrho_{N,t}^{(\Lambda\cup \{k_1,\dots k_m\})}$ with $t<t_0$, the convergence of the series can be iteratively extended to arbitrary times. To prove the theorem, we will proceed in two steps. The first step consists in showing that, looking at Eq.~\eqref{eq:formal_sol}, in the term $\mathcal{L}^{(\Lambda)}$ the interaction terms can be neglected in the thermodynamic limit. In the second step, we show instead that the sum over $k$ in Eq.~\eqref{eq:formal_sol} can be extended to a sum over all sites but the $j$th one. As we discuss, this implies the statement of the theorem. \\

\noindent {\bf First step.} To complete the first step, we show that series in Eq.~\eqref{series_T} is, in the thermodynamic limit, equivalent to the same series with  $\mathcal{L}^{(\Lambda)}$ substituted by
$$
\hat{\mathcal{L}}^{(\Lambda)}=\sum_{k\in \Lambda}\mathcal{L}_k^{(k)}\, .
$$
This is true in the regime $\alpha\le d$, considered in the theorem. To show this, we proceed as follows. We construct the state $\hat{\varrho}_{N,t}^{(\Lambda)}$ as 
$$
\hat{\varrho}_{N,t}^{(\Lambda)}=\sum_{m=0}^{N-|\Lambda|} \int_0^{t}{\rm d}t_1\int_0^{t_1}{\rm d}t_2\dots \int_0^{t_{m-1}}{\rm d}t_m \, \hat{T}_{m}\, ,
$$
where $\hat{T}_m$ are analogous to the terms in Eq.~\eqref{series_term}, with $\hat{\Phi}_m$ as in Eq.~\eqref{series_term2} but with $\mathcal{L}^{(\Lambda\cup\{k_1, \dots k_n\})}$ substituted by $\hat{\mathcal{L}}^{(\Lambda\cup\{k_1,\dots k_n\})}$. Following the same steps presented in Lemma~\ref{lemma1}, also such a series can be shown to converge, uniformly in $N$, for all times. 

We consider the difference for $t<t_0$, in trace norm, between the two density matrices represented via the series. We want to show that 
\begin{equation}
\lim_{N\to\infty}\left\|\varrho_{N,t}^{(\Lambda)}-\hat{\varrho}_{N,t}^{(\Lambda)}\right\|_{\rm tr}=0\, .
\label{lim_step_1}
\end{equation}
To this end, we consider the following bound 
$$
\left\|\varrho_{N,t}^{(\Lambda)}-\hat{\varrho}_{N,t}^{(\Lambda)}\right\|_{\rm tr}\le \|S_M-\hat{S}_M\|_{\rm tr}+\|R_M\|_{\rm tr}+\|\hat{R}_M\|_{\rm tr}\, ,
$$
where we have introduced the truncated series 
$$
S_M=\sum_{m=0}^{M} \int_0^{t}{\rm d}t_1\int_0^{t_1}{\rm d}t_2\dots \int_0^{t_{m-1}}{\rm d}t_m \, T_{m}\, , \qquad \hat{S}_M=\sum_{m=0}^{M} \int_0^{t}{\rm d}t_1\int_0^{t_1}{\rm d}t_2\dots \int_0^{t_{m-1}}{\rm d}t_m \, \hat{T}_{m}\, , 
$$
with $M<N-|\Lambda|$. The terms  $R_M=\varrho_{N,t}^{(\Lambda)}-S_M$ and $\hat{R}_M=\hat{\varrho}_{N,t}^{(\Lambda)}-\hat{S}_M$  denote the remainders of the two series. To show that the difference is vanishingly small in the large $N$ limit, we  exploit an $\epsilon/3$-argument. 
The trace-norm convergence of both perturbation series for $t<t_0$ allows us to make the terms $\|R_M\|_{\rm tr}$ and $\|\hat{R}_M\|_{\rm tr}$ smaller than $\epsilon/3$, independently of $N$, for any $\epsilon>0$, by choosing a sufficiently large $M$. 
It thus remains to be shown that the difference $\|S_M-\hat{S}_M\|_{\rm tr}$ can also be made smaller than $\epsilon/3$, for any $\epsilon>0$,  for large enough $N$. The terms of the two series solely differ by the fact that in $\hat{T}_m$ the evolution in between collisions is implemented by $\hat{\mathcal{L}}^{(\Delta)}$ while in $T_m$ it is implemented by  $\mathcal{L}^{(\Delta)}$. Here, $\Delta$ denotes a generic finite set of sites. Due to Lemma \ref{lemma2}, we know that $e^{t\hat{\mathcal{L}}^{(\Delta)}}\to e^{t \mathcal{L}^{(\Delta)}}$, in the thermodynamic limit and in trace norm, whenever the two maps act on trace-class operators. This means that it is always possible to find $N$ large enough, such that $\|S_M-\hat{S}_M\|_{\rm tr}\le \epsilon/3$, which thus implies the validity of Eq.~\eqref{lim_step_1}.

To extend the equivalence of the perturbation series for arbitrary times, we proceed by induction as follows. Let us assume that for $t=t^*$ we have that the limit in Eq.~\eqref{lim_step_1} holds. We then write the states at any time $t$ after $t^*$ through a perturbation series evolving the states from $t^*$ onward as
\begin{equation}
\varrho_{N,t,t^*}^{(\Lambda)}=\sum_{m=0}^{N-|\Lambda|} \int_{t^*}^{t}{\rm d}t_1\int_{t^*}^{t_1}{\rm d}t_2\dots \int_{t^*}^{t_{m-1}}{\rm d}t_m \, \sum_{j_1,..j_m,k_1,...k_m}\Phi_m[\varrho_{N,t^*}^{(\Lambda\cup\{k_1,...k_m\})}]\, ,
\label{extension_1}
\end{equation}
and 
$$
\hat{\varrho}_{N,t,t^*}^{(\Lambda)}=\sum_{m=0}^{N-|\Lambda|} \int_{t^*}^{t}{\rm d}t_1\int_{t^*}^{t_1}{\rm d}t_2\dots \int_{t^*}^{t_{m-1}}{\rm d}t_m \, \sum_{j_1,..j_m,k_1,...k_m}\hat{\Phi}_m[\hat{\varrho}_{N,t^*}^{(\Lambda\cup\{k_1,...k_m\})}]\, .
$$
In these equations, the sums run over the indexes as explicitly written in Eq.~\eqref{series_term}. Moreover, because of Lemma \ref{lemma1}, the above series converge for $t-t^*<t_0$. As a consequence, recalling the $\epsilon/3$ argument exploited above, we can ensure that the two series are equivalent, for large $N$,   by simply showing that 
$$
\lim_{N\to\infty}\left\|{\Phi}_m\left[{\varrho}_{N,t^*}^{(\Lambda\cup\{k_1,...k_m\})}\right]-\hat{\Phi}_m\left[\hat{\varrho}_{N,t^*}^{(\Lambda\cup\{k_1,...k_m\})}\right]\right\|_{\rm tr}=0\, .
$$
To this end, we write 
\begin{equation}
\begin{split}
\left\|{\Phi}_m\left[{\varrho}_{N,t^*}^{(\Lambda\cup\{k_1,...k_m\})}\right]-\hat{\Phi}_m\left[\hat{\varrho}_{N,t^*}^{(\Lambda\cup\{k_1,...k_m\})}\right]\right\|_{\rm tr}&\le \left\|{\Phi}_m\left[{\varrho}_{N,t^*}^{(\Lambda\cup\{k_1,...k_m\})}-\hat{\varrho}_{N,t^*}^{(\Lambda\cup\{k_1,...k_m\})}\right]\right\|_{\rm tr}\\
&+\left\|{\Phi}_m\left[\hat{\varrho}_{N,t^*}^{(\Lambda\cup\{k_1,...k_m\})}\right]-\hat{\Phi}_m\left[\hat{\varrho}_{N,t^*}^{(\Lambda\cup\{k_1,...k_m\})}\right]\right\|_{\rm tr}\, .
\end{split}
\label{}
\end{equation}
The first term on the right hand side of the inequality vanishes, in the thermodynamic limit, due to our assumption on the equivalence of the series for $t=t^*$. The second term on the right hand side vanishes due to the result of Lemma \ref{lemma2}, as discussed before. Therefore, starting from $t^*<t_0$ and using the induction step presented above we can extended the validity of the limit in Eq.~\eqref{lim_step_1} to arbitrary times. \\

\noindent {\bf Second step.} In the second step of the proof, we construct a new family of reduced states and show that this is equivalent to the family associated with the states  $\hat{\varrho}_{N,t}^{(\Lambda)}$, for large $N$. More precisely, we consider reduced states defined as 
$$
\tilde{\varrho}_{N,t}^{(\Lambda)}=\bigotimes_{j\in\Lambda} \tilde{\varrho}_{N,t}^{(j)}\, ,
$$
with single-site reduced states obeying the differential equation reported in Eq.~\eqref{eq:effective_generator} of the main text. This implies that the state $\tilde{\varrho}_{N,t}^{(\Lambda)}$ obeys the equation 
\begin{equation}
    \dot{ \tilde{\varrho}}_{N,t}^{(\Lambda)} = \hat{\mathcal{L}}^{(\Lambda)}\left[ \tilde{\varrho}_{N,t}^{(\Lambda)} \right] + \frac{1}{c_\alpha^N}\underset{j\in\Lambda}{\sum} \sum_{\substack{k\in\Gamma\\ k\neq j}}\frac{\mathrm{tr}_{k} \left( \mathcal{L}_{jk}^{(jk)}\left[ \tilde{\varrho}_{N,t}^{(\Lambda\cup \{k\})}\right]\right)}{||\textbf{r}_j-\textbf{r}_k||_\Gamma^\alpha}\, ,
    \label{eq_hat}
\end{equation}
which only differs from the one of $\hat{\varrho}_{N,t}^{(\Lambda)}$ since the summation over $k$ also includes sites within the set $\Lambda$. Note that, following the approach of Ref.~\cite{benedikter2016}, it can be shown that the solution of Eq.~\eqref{eq_hat} is unique and thus given by  reduced product states whenever the initial state is in product form. 
The equivalence between $\hat{\varrho}_{N,t}^{(\Lambda)}$ and $\tilde{\varrho}_{N,t}^{(\Lambda)}$, in the thermodynamic limit, can be shown investigating the corresponding perturbation series. Through the same procedure exploited above, we can construct a perturbation series for $\tilde{\varrho}_{N,t}^{(\Lambda)}$ from the differential equation in Eq.~\eqref{eq_hat}. The term $\tilde{T}_m$ of the series [cf.~the analogous terms in Eq.~\eqref{series_term}] only differ from $\hat{T}_m$ for the different summations over the indexes $k_i$'s. Due to Lemma \ref{lemma1}, also this series converges initially for $t<t_0$ but can be extended to arbitrary times.  

Considering the difference between the perturbation series for $\hat{\varrho}_{N,t}^{(\Lambda)}$ and the one for $\tilde{\varrho}_{N,t}^{(\Lambda)}$ and exploiting their convergence, we can again use an $\epsilon/3$ argument to reduce the task of showing that 
\begin{equation}
\lim_{N\to\infty}\|\hat{\varrho}_{N,t}^{(\Lambda)}-\tilde{\varrho}_{N,t}^{(\Lambda)}\|_{\rm tr}=0
\label{lim_step2}
\end{equation}
to the one of showing 
$$
\lim_{N\to\infty}\|\hat{T}_m -\tilde{T}_m\|_{\rm tr}=0\, .
$$
The terms $\hat{T}_m$ and  $\tilde{T}_m$ only differ  for the fact that the summations over $k_i's$ in $\tilde{T}_m$ include more terms. Splitting such summations into two parts, one equivalent to the one in $\hat{T}_m$ plus the additional terms, gives 
\begin{align*}
    \tilde{T}_m&=\frac{1}{(c_\alpha^N)^m}\!\sum_{j_1\in\Lambda}\!\left(\sum_{k_1\notin \Lambda}+\sum_{\substack{k_1\in\Lambda\\k_1\neq j_1}}\right)\dots\!\!\! \sum_{j_m\in \Lambda\cup \{k_1,k_2,\dots k_{m-1}\}}\!\left(\sum_{k_m\notin \Lambda\cup \{k_1,k_2,\dots k_{m-1}\}}\!+\!\sum_{\substack{k_m\in \Lambda\cup \{k_1,k_2,\dots k_{m-1}\}\\k_m \neq j_m}} \right)\!\hat{\Phi}_m[\varrho_{N,0}^{(\Lambda\cup \{k_1,\dots k_m\})}] \, . 
\end{align*}
Expanding the above terms by taking the products between different sums over $k_i$ we find that 
$$
\tilde{T}_m=\hat{T}_m+Z_m\, ,
$$
where $Z_m$ is a correction consisting of $2^m-1$ bounded terms. In all of these terms, there appears at least one sum over $k_i$ which is not extensive in $N$ (indeed the term in which all sums are extensive in $N$ is $\hat{T}_m$). Since all sums are associated with a factor $1/c_\alpha^N$, the nonextensive summation makes it such that, at leading order, the correction $Z_m$ scales as
$$
\|Z_m\|_{\rm tr}\sim \frac{1}{c_\alpha^N}\, ,
$$
which  results in $\lim_{N\to\infty}\|\hat{T}_m-\tilde{T}_m\|_{\rm tr}\to 0$. As done at the end of the first part of the proof, exploiting the argument above we can prove the validity of the limit in Eq.~\eqref{lim_step2} first for $t<t_0$ and then extend it to arbitrary times. 
\qed

\newpage

\section{II. Proof of the proposition}
In this section, we provide the proof of the proposition stated in the main text. \\

\noindent {\it Proof:} To study convergence of reduced states in the thermodynamic limit, we introduce the rescaled coordinate $\textbf{x}_j = \textbf{r}_j/L$, with entries given by $\textbf{x}_j^\mu=1/L,2/L, \dots 1$. In the limit $N\to\infty$, each site is thus identified through its (continuous) coarse-grained position in the infinite lattice. With such a coordinate, we denote reduced single-particle states as $\tilde{\varrho}_{N,t}^{(j)}=\tilde{\varrho}_{N,t}^{(\textbf{x}_j )}$. From Theorem \ref{theorem}, the perturbation series for $\tilde{\varrho}_{N,t}^{(\textbf{x}_j)}$ [see discussion after  Eq.~\eqref{eq_hat}] converges uniformly in $N$ for $t<t_0$. As such, the task of determining the limiting expression $ \tilde{\varrho}_{\infty,t}^{(\textbf{x})} = \underset{N\to\infty}{\lim} \tilde{\varrho}_{N,t}^{(\textbf{x}_j)}$, with $\lim_{N\to\infty}\textbf{x}_j=\textbf{x}$, reduces to controlling the limit for the terms $ \tilde{T}_m$. To this end, it is sufficient to show that the sums over $\textbf{y}_{k_i}$'s in $\tilde{T}_m$ [see  Eq.~\eqref{eq:tilde_T_m} below] define Riemann sums and converge, in the thermodynamic limit, to integrals. 

Recalling the scaling of the Kac factor for $\alpha<d$ and $N\gg 1$ [cf. Eq.~\eqref{Kac}], we find
\begin{equation} \label{eq:tilde_T_m}
\begin{split}
\tilde{T}_m \approx \theta^m(\alpha) \frac{1}{N}\sum_{\textbf{y}_{k_1}\neq \textbf{x}_{j}}\sum_{\textbf{x}_{j_2}\in\{ \textbf{x}_{j},\textbf{y}_{k_1}\}}\frac{1}{N}\sum_{\textbf{y}_{k_2}\neq \textbf{x}_{j_2}}\dots & \sum_{\textbf{x}_{j_m}\in\{ \textbf{x}_{j},\textbf{y}_{k_1},\dots\textbf{y}_{k_{m-1}}\}}\frac{1}{N}\sum_{\textbf{y}_{k_m}\neq \textbf{x}_{j_m}}\\
&\times \frac{\tilde{\Phi}_m'\left[ \tilde{\varrho}_{N,0}^{(\textbf{x}_{j})}\otimes\tilde{\varrho}_{N,0}^{(\textbf{y}_{k_1})}\otimes\dots\otimes \tilde{\varrho}_{N,0}^{(\textbf{y}_{k_m})}\right]}{\|{\bf x}_{j}-{\bf y}_{k_1}\|_\Gamma^\alpha \|{\bf x}_{j_2}-{\bf y}_{k_2}\|_\Gamma^\alpha \dots \|{\bf x}_{j_m}-{\bf y}_{k_m}\|_\Gamma^\alpha}\, ,
\end{split}
\end{equation}
with 
\begin{equation} \label{eq:tilde_phi_m}
\begin{split}
    \tilde{\Phi}_m'\left[ \tilde{\varrho}_{N,0}^{(\textbf{x}_{j})}\otimes\tilde{\varrho}_{N,0}^{(\textbf{y}_{k_1})}\otimes\dots\otimes \tilde{\varrho}_{N,0}^{(\textbf{y}_{k_m})}\right]&=e^{(t-t_1)\hat{\mathcal{L}}^{(\{\textbf{x}_{j}\})}}\circ{\rm tr}_{\textbf{y}_{k_1}}\circ\mathcal{L}_{12}^{(\textbf{x}_{j} \textbf{y}_{k_1})}\circ\dots\\
    &\circ {\rm tr}_{\textbf{y}_{k_m}}\circ \mathcal{L}_{12}^{(\textbf{x}_{j_m}\textbf{y}_{k_m} )} \circ e^{t_m\hat{\mathcal{L}}^{(\{\textbf{x}_{j},\textbf{y}_{k_1},\dots\textbf{y}_{k_m}\})}}\left[ \tilde{\varrho}_{N,0}^{(\textbf{x}_{j})}\otimes\tilde{\varrho}_{N,0}^{(\textbf{y}_{k_1})}\otimes\dots\otimes \tilde{\varrho}_{N,0}^{(\textbf{y}_{k_m})}\right]\, .
    \end{split}
\end{equation}
When solely considering the inverse power-law functions with $\alpha<d$, the (Riemann) sums would converge to improper integrals. To make sure that such a converge still occurs when including the operator $\tilde{\Phi}_m'\left[ \tilde{\varrho}_{N,0}^{(\textbf{x}_{j})}\otimes\tilde{\varrho}_{N,0}^{(\textbf{y}_{k_1})}\otimes\dots\otimes \tilde{\varrho}_{N,0}^{(\textbf{y}_{k_m})}\right]$, we need to show that the latter is continuous function, in trace norm and in the thermodynamic limit, of the coordinates $\textbf{x}_{j},\textbf{y}_{k_1},\dots\textbf{y}_{k_m}$, if the initial single-particle states are. To this end, we note that the superscripts associated with the homogeneous interaction pairs, e.g.  $\mathcal{L}_{12}^{(\textbf{x}_{j} \textbf{y}_{k_1})}$, solely indicate on which entries of the tensor product the generators are acting nontrivially. The latter generators do not otherwise explicitly depend on the rescaled  coordinates. With this observation, it can be shown that 
\begin{equation}\label{cont-phi}
\begin{split}
    \Big\|\tilde{\Phi}_m'\left[\tilde{\varrho}_{N,0}^{(\textbf{x}_{j})}\otimes\tilde{\varrho}_{N,0}^{(\textbf{y}_{k_1})}\otimes\dots\otimes \tilde{\varrho}_{N,0}^{(\textbf{y}_{k_m})}\right] &- \tilde{\Phi}_m'\left[ \tilde{\varrho}_{N,0}^{(\textbf{x}_{j})}\otimes\tilde{\varrho}_{N,0}^{(\textbf{z}_{k_1})}\otimes\dots\otimes \tilde{\varrho}_{N,0}^{(\textbf{z}_{k_m})}\right]\Big\|_{\mathrm{tr}}\\
    &\le C^m\left\|\tilde{\varrho}_{N,0}^{(\textbf{x}_{j})}\otimes\tilde{\varrho}_{N,0}^{(\textbf{y}_{k_1})}\otimes\dots\otimes \tilde{\varrho}_{N,0}^{(\textbf{y}_{k_m})}-  \tilde{\varrho}_{N,0}^{(\textbf{x}_{j})}\otimes\tilde{\varrho}_{N,0}^{(\textbf{z}_{k_1})}\otimes\dots\otimes \tilde{\varrho}_{N,0}^{(\textbf{z}_{k_m})} \right\|_{\mathrm{tr}} \, ,
\end{split}
\end{equation}
where $C$ is given in Lemma~\ref{lemma1}.
By assumption the initial reduced single-particle states $\tilde{\varrho}_{N,0}^{(\textbf{y}_{k_i})}$ converge, in trace norm, to the states $\tilde{\varrho}_{\infty,0}^{(\textbf{y}_{k_i})}$ which are continuous in the re-scaled coordinates. This fact, together with Eq.~\eqref{cont-phi}, guarantees the continuity of the operator $\tilde{\Phi}_m'\left[ \tilde{\varrho}_{N,0}^{(\textbf{x}_{j})}\otimes\tilde{\varrho}_{N,0}^{(\textbf{y}_{k_1})}\otimes\dots\otimes \tilde{\varrho}_{N,0}^{(\textbf{y}_{k_m})}\right]$, in the thermodynamic limit. 

Exploiting such a continuity, all sums over $\textbf{y}_{k_i}$'s converge to integrals in the thermodynamic limit, giving  
\begin{equation}
\begin{split}
\label{T_tilde_cont}
    \underset{N\to\infty}{\lim} \tilde{T}_m =\theta^m(\alpha)\!\!\! \sum_{\textbf{x}_2,\dots \textbf{x}_m} &\int {\rm d}\textbf{y}_1 \dots \int {\rm d}\textbf{y}_m\\
    &\times \frac{e^{(t-t_1)\hat{\mathcal{L}}^{(\textbf{x})}}\circ{\rm tr}_{\textbf{y}_1}\circ\mathcal{L}_{12}^{(\textbf{x}\textbf{y}_1)}\dots{\rm tr}_{\textbf{y}_m}\circ \mathcal{L}_{12}^{(\textbf{x}_m\textbf{y}_m)} \circ e^{t_m\hat{\mathcal{L}}^{(\{\textbf{x},\textbf{y}_1,\dots \textbf{y}_m\})}}\left[ \tilde{\varrho}_{\infty,0}^{(\textbf{x})}\otimes\dots\otimes \tilde{\varrho}_{\infty,0}^{(\textbf{y}_m)}\right]}{\|{\bf x}-{\bf y}_{1}\|_\Gamma^\alpha\|{\bf x}_{2}-{\bf y}_{2}\|_\Gamma^\alpha \dots \|{\bf x}_{m}-{\bf y}_{m}\|_\Gamma^\alpha} \, .
    \end{split}
\end{equation}
Each element of the series is continuous in the coordinate $\textbf{x}$ and the series uniformly converges to the limiting state $\tilde{\varrho}_{\infty,t}^{(\textbf{x})}$, for $t<t_0$, which is thus also continuous in $\textbf{x}$. What is left to show is that the perturbation series in the thermodynamic limit is the (unique) solution of Eq.~\eqref{eq:dyn_proposition}. 
This can be argued as follows. 
In addition to $\tilde{\varrho}_{N,t}^{(\textbf{x}_j)}$ also its derivative $\dot{\tilde{\varrho}}_{N,t}^{(\textbf{x}_j)}$ [cf.~Eq.~\eqref{eq:effective_generator}] clearly converges uniformly in $N$ for $t\le t_0$ such that $\underset{N\to\infty}{\lim} \dot{\tilde{\varrho}}_{N,t}^{(\textbf{x}_j)} = \dot{\tilde{\varrho}}_{\infty,t}^{(\textbf{x})}$. Using the latter relation as well as $\underset{N\to\infty}{\lim} {\tilde{\varrho}}_{N,t}^{(\textbf{x}_j)} = {\tilde{\varrho}}_{\infty,t}^{(\textbf{x})}$ on both sides of Eq.~\eqref{eq:effective_generator} and exploiting the continuity in $\textbf{x}$, the sum over the collisions in Eq.~\eqref{eq:effective_generator}
can be shown to converge to the improper integral in Eq.~\eqref{eq:dyn_proposition}. This demonstrates that the limit of the reduced states $\varrho_{N,t}^{(\textbf{x}_j)}$ converge to the solution of  Eq.~\eqref{eq:dyn_proposition} [see also Lemma \ref{lemma3} for an alternative proof of the continuity of the solution of Eq.~\eqref{eq:dyn_proposition}]. 

We note that the same convergence result could be shown also by considering reduced states $\varrho_{N,t}^{(\Lambda)}$ over finite sets of sites $\Lambda$. In this case, using the arguments above for the convergence of the corresponding terms $\tilde{T}_m$, one can show that $\varrho_{\infty,t}^{(\Lambda)}$ solves an equation analogous to Eq.~\eqref{eq_hat} but with collision term given by the integrals. Following the arguments of Ref.~\cite{benedikter2016}, the latter equation has a unique solution which is given by the tensor-product states. Finally, using this information gives rise to Eq.~\eqref{eq:dyn_proposition} for the reduced states $\varrho_{\infty,t}^{(\textbf{x})}$. \\

As done in the proof of the main theorem, we can extend the convergence of the reduced state $\tilde{\varrho}_{N,t}^{(\textbf{x}_j)}$ to the solution of Eq.~\eqref{eq:dyn_proposition}, at all times, by induction. The reasoning is as follows. For a generic induction step we assume that there exists a $t^*$ such that, in trace norm, 
$$
\lim_{N\to\infty}\tilde{\varrho}_{N,t^*}^{(\textbf{x}_j)}=\tilde{\varrho}_{\infty,t^*}^{(\textbf{x})}
$$
where the single-particle states are continuous in the rescaled coordinate and solve Eq.~\eqref{eq:dyn_proposition}. Then, we can define the perturbation series solution of Eq.~\eqref{eq_hat} starting from $t^*$. This reads as 
\begin{equation}
    \tilde{\varrho}_{N,t,t^*}^{(\textbf{x}_j)}=\sum_{m=0}^{N-1} \int_{t^*}^{t}{\rm d}t_1\int_{t^*}^{t_1}{\rm d}t_2\dots \int_{t^*}^{t_{m-1}}{\rm d}t_m \, \sum_{j_1,..j_m,k_1,...k_m}\frac{\tilde{\Phi}_m'\left[ \tilde{\varrho}_{N,t^*}^{(\textbf{x}_{j})}\otimes\tilde{\varrho}_{N,t^*}^{(\textbf{y}_{k_1})}\otimes\dots\otimes \tilde{\varrho}_{N,t^*}^{(\textbf{y}_{k_m})}\right]}{\|{\bf x}_{j}-{\bf y}_{k_1}\|_\Gamma^\alpha \|{\bf x}_{j_2}-{\bf y}_{k_2}\|_\Gamma^\alpha \dots \|{\bf x}_{j_m}-{\bf y}_{k_m}\|_\Gamma^\alpha}\, .
\label{ind-assum}
\end{equation}
Due to the argument above, the perturbation series converges uniformly to the continuous reduced state $\tilde{\varrho}_{\infty,t}^{(\textbf{x})}$ solution of Eq.~\eqref{eq:dyn_proposition} within the time window $[t^*,t^*+t_0)$.\\
For the first induction step, we can thus take  $t^*<t_0$ for which the previous part of the proof guarantees the induction assumption in Eq.~\eqref{ind-assum}. Then, exploiting the same reasoning which lead to Eq.~\eqref{T_tilde_cont} we can extend the solution to the time $t^{**}\in[t^*,t^*+t_0)$. Iterating the above induction step we can show convergence to the solution at all times.

\newpage

\section{III. Lemmata}
\begin{lemma}\label{lemma1}
The perturbation-series expansion for $\varrho_{N,t}^{(\Lambda)}$ with terms as in Eq.~\eqref{series_term} converges, in trace norm and  uniformly in $N$, for $t<t_0=2^d C$. Here, $C$ is the $N$-independent constant 
\begin{equation}
    \label{C-const}
    C=\sup_{\substack{j,k\in\Gamma\\j\neq k}}\left(2 ||V_{jk}^{(jk)}||+ \sum_\mu |w_{jk}^\mu|\,   ||L^{(j)}_\mu||\, ||L_\mu^{(k)\dagger}||+\sum_\mu |w_{kj}^\mu|\,  ||L^{(k)}_\mu||\, ||L_\mu^{(j)\dagger}||\, \right)\, .
\end{equation}
\end{lemma}
\noindent\textit{Proof.} We start by showing that each term $T_m$ [cf.~Eq.~\eqref{series_term}] of the perturbation series is uniformly bounded in trace norm. To this end, we note that given any completely positive and trace-preserving map $\Psi$, we have $\|\Psi[X]\|_{\rm tr}\le \|X\|_{\rm tr}$. 
Moreover, considering our dynamical generators $\mathcal{L}_{jk}^{(jk)}$, we also have 
$\left\|\mathcal{L}^{(jk)}_{jk}[X] \right\|_{\mathrm{tr}} \le C \left\| X \right\|_{\mathrm{tr}}$. 
This relation holds for a generic trace-class operator $X$ and we exploits that  $\left\|A B \right\|_{\mathrm{tr}} \le  \|B\| \left\| A \right\|_{\mathrm{tr}}$, for any bounded operator $B$ and any  trace-class operator $A$, where $\|\cdot\|$ denotes the  operator norm.
Using these bounds, we can  show that (noticing that also the partial trace is a completely positive and trace-preserving map)
$$
    \left\|\mathcal{L}_{j_1k_1}^{(j_1k_1)}\circ e^{(t_1-t_2)\mathcal{L}^{(\Lambda\cup \{k_1\})}}\circ \dots \circ {\rm tr}_{k_m}\circ \mathcal{L}_{j_mk_m}^{(j_mk_m)} \circ e^{t_m\mathcal{L}^{(\Lambda\cup \{k_1,\dots k_m\})}}\left[\varrho_{N,0}^{(\Lambda\cup \{k_1,\dots k_m\})}\right]\right\|_{\mathrm{tr}}\le C^m \, ,
$$
which implies 
$$
    \left\|T_m^N\right\|_{\mathrm{tr}}\le\left(\frac{C}{c_\alpha^N}\right)^m\sum_{j_1\in\Lambda}\sum_{k_1\notin\Lambda}\dots \sum_{j_m\in \Lambda\cup \{k_1,k_2,\dots k_{m-1}\}}\sum_{k_m\notin \Lambda\cup \{k_1,k_2,\dots k_{m-1}\}}\frac{1}{\|{\bf r}_{j_1}-{\bf r}_{k_1}\|^\alpha_\Gamma \dots \|{\bf r}_{j_m}-{\bf r}_{k_m}\|^\alpha_\Gamma}\, .
$$
To give a bound for the expression above, we extend the sums over all $k_i$s to all sites but with $k_i\neq j_i$. We have that 
\begin{align*}
  \sum_{k_i\notin \Lambda\cup \{k_1,k_2,\dots k_{i-1}\}} \frac{1}{||\textbf{r}_{j_i}-\textbf{r}_{k_i}||_\Gamma^\alpha} &< \sum_{\substack{k_i \in\Gamma\\ k_i\neq j_i}} \frac{1}{||\textbf{r}_{j_i}-\textbf{r}_{k_i}||_\Gamma^\alpha} \leq 2^d c_\alpha^N\, .
\end{align*}
We note that the last sum can be rewritten as $\sum_{\Delta\textbf{r}\neq 0} ||\Delta\textbf{r}||_\Gamma^{-\alpha}$, with $\Delta\textbf{r} = \textbf{r}_{j_i}-\textbf{r}_{k_i}$, which is similar to the definition of the Kac factor but with different summation indexes. In each dimension the summation indeed splits into the two parts $\textbf{r}_{j_i}^\mu-1,\dots, 0$ and $-1,\dots, \textbf{r}_{j_i}^\mu-L$ and by adding terms to both parts in every dimension one finally recovers $2^d$ times the Kac factor. Moreover, each sum over $j_i$ is bounded by $n+i-1$, where $n=|\Lambda|$ is the extension of the subset $\Lambda$. All these considerations give  
$$
\left\|\int_0^{t}{\rm d}t_1\int_0^{t_1}{\rm d}t_2\dots \int_0^{t_{m-1}}{\rm d}t_m \, T_{m} \right\|_{\rm tr}\le \frac{(2^d C t)^m }{m!}n(n+1)(n+2)\dots (n+m-1)\, ,
$$
and thus the series converges for $t< t_0=2^dC$.  \qed 
\\
\begin{lemma}\label{lemma2}
Consider the dynamical generator $\mathcal{L}^{(\Lambda)}$ given in Eq.~\eqref{eq:gen_lambda} and the generator $\hat{\mathcal{L}}^{(\Lambda)}=\sum_{k\in \Lambda}\mathcal{L}_k^{(k)}$, with $\mathcal{L}_k^{(k)}$ given in Eq.~\eqref{L1}. For $\alpha\le d$, we have 
\begin{equation}\label{eq:convergence_eff_evol}
    \underset{N\to\infty}{\lim} \left\|e^{t\mathcal{L}^{(\Lambda)}} [Q^{(\Lambda)}]- e^{t\hat{\mathcal{L}}^{(\Lambda)}} [Q^{(\Lambda)}]\right\|_{\mathrm{tr}} =  0 \, ,
\end{equation}
for any trace-class operator $Q^{(\Lambda)}$ defined on the sites in $\Lambda$.
\end{lemma}
\noindent \textit{Proof.} We follow and adapt the approach of Lemma 1 in Ref.~\cite{carollo2023nongaussian}. As a starting point, we note that the difference between the two evolutions can be written as 
\begin{equation*}
\begin{split}
    e^{t\mathcal{L}^{(\Lambda)}} [Q^{(\Lambda)}]- e^{t\hat{\mathcal{L}}^{(\Lambda)}} [Q^{(\Lambda)}]&=\int_0^t {\rm d}{s} \frac{\rm d}{{\rm d}s}  e^{s\mathcal{L}^{(\Lambda)}} \circ e^{(t-s)\hat{\mathcal{L}}^{(\Lambda)} } [Q^{(\Lambda)}]=\int_0^t{\rm d}s \, e^{s\mathcal{L}^{(\Lambda)}}\circ \left(\mathcal{L}^{(\Lambda)}-\hat{\mathcal{L}}^{(\Lambda)}\right)\circ e^{(t-s)\hat{\mathcal{L}}^{(\Lambda)}}\left[Q^{(\Lambda)}\right]\, .
    \end{split}
\end{equation*}
Taking the trace norm of the integrand and considering that the generators have the same local terms, we find the bound 
\begin{align*}
    \left\| e^{s\mathcal{L}^{(\Lambda)}} \circ \left[\mathcal{L}^{(\Lambda)}- \hat{\mathcal{L}}^{(\Lambda)}\right] \circ e^{(t-s)\hat{\mathcal{L}}^{(\Lambda)}} [Q^{(\Lambda)}] \right\|_{\rm tr}
    &\leq \left\| \frac{1}{2c_\alpha^N} \sum_{j,k\in \Lambda} \frac{\mathcal{L}_{jk}^{(jk)}\circ e^{(t-s)\hat{\mathcal{L}}^{(\Lambda)}} [Q^{(\Lambda)}]}{[\delta_{jk}+\|\textbf{r}_j-\textbf{r}_k\|^\alpha_\Gamma]}\right\|_{\rm tr}
    \leq \left\| Q^{(\Lambda)} \right\|_{\rm tr} \frac{C|\Lambda|^2}{c_\alpha^N}\, .
\end{align*}
Recalling the scaling of the Kac factor given in Eq.~\eqref{Kac}, for $\alpha\le d$, we thus have 
\[\underset{N\to\infty}{\lim} \left\|e^{t\mathcal{L}^{(\Lambda)} } [Q^{(\Lambda)}]- e^{t \hat{\mathcal{L}}^{(\Lambda)} } [Q^{(\Lambda)}]\right\|_{\rm tr} = \lim_{N\to\infty} \left\| Q^{(\Lambda)} \right\|_{\rm tr} \frac{C|\Lambda|^2t}{c_\alpha^N}=0
\]
for any finite set $\Lambda$ and  arbitrary time  $t$.\qed \\

\begin{lemma}\label{lemma3}
     Assuming initial single-site reduced states $\tilde{\varrho}_{\infty,0}^{(\bf{x})}$ which are continuous, in trace norm, in the rescaled coordinate $\bf{x}$, the reduced states $\tilde{\varrho}_{\infty,t}^{(\bf{x})}$ solving  Eq.~\eqref{eq:dyn_proposition} remain continuous at all times. 
\end{lemma}
\noindent \textit{Proof.} We consider the difference $\Delta_{\infty, t}^{\textbf{x}\textbf{y}} = \tilde{\varrho}_{\infty,t}^{(\textbf{x})} - \tilde{\varrho}_{\infty,t}^{(\textbf{y})}$, which evolves according to
\begin{align*}
    \dot{\Delta}_{\infty,t}^{(\textbf{x}\textbf{y})} &=  \mathcal{L}_1\left[  \tilde{\varrho}_{\infty,t}^{(\textbf{x})}\right] + \int {\rm d}\textbf{z} \frac{\mathrm{tr}_{\textbf{z}} \circ \mathcal{L}_{12}\circ\left[  \tilde{\varrho}_{\infty,t}^{(\textbf{x})} \otimes \tilde{\varrho}_{\infty,t}^{(\textbf{z})}\right] }{\|\textbf{x}-\textbf{z}\|_\Gamma^\alpha} -  \mathcal{L}_1\left[  \tilde{\varrho}_{\infty,t}^{(\textbf{y})}\right] - \int {\rm d}\textbf{z} \frac{\mathrm{tr}_{\textbf{z}} \circ \mathcal{L}_{12}\circ\left[  \tilde{\varrho}_{\infty,t}^{(\textbf{y})} \otimes \tilde{\varrho}_{\infty,t}^{(\textbf{z})}\right] }{\|\textbf{y}-\textbf{z}\|_\Gamma^\alpha}\\\
    &=\mathcal{L}_1\left[ \Delta_{\infty,t}^{\textbf{x}\textbf{y}}\right] + \int {\rm d}\textbf{z} \frac{\mathrm{tr}_{\textbf{z}} \circ \mathcal{L}_{12}\circ\left[ \Delta_{\infty,t}^{\textbf{x}\textbf{y}} \otimes \tilde{\varrho}_{\infty,t}^{(\textbf{z})}\right] }{\|\textbf{x}-\textbf{z}\|_\Gamma^\alpha} + \left(\int {\rm d}\textbf{z} \frac{1}{\|\textbf{x}-\textbf{z}\|_\Gamma^\alpha} -\int {\rm d}\textbf{z} \frac{1}{\|\textbf{y}-\textbf{z}\|_\Gamma^\alpha}\right)\mathrm{tr}_{\textbf{z}} \circ \mathcal{L}_{12}\circ\left[  \tilde{\varrho}_{\infty,t}^{(\textbf{y})} \otimes \tilde{\varrho}_{\infty,t}^{(\textbf{z})}\right]\\\
    &= \mathcal{M}_t\left[ \Delta_{\infty,t}^{\textbf{x}\textbf{y}}\right] + \mathcal{A}_t\, .
\end{align*}
At time $t$ the difference is given by the formal solution $ \Delta_{\infty,t}^{(\textbf{x}\textbf{y})} = \mathcal{V}_{t,0}[\Delta_{\infty,0}^{(\textbf{x}\textbf{y})}] + \int_0^t {\rm d}s \mathcal{V}_{t,s}[\mathcal{A}_s]$, with the propagator
$
    \mathcal{V}_{t,s} = \overset{\leftarrow}{\mathcal{T}}\exp(\int_s^t {\rm d}u \mathcal{M}_u)\, ,
$
where $\overset{\leftarrow}{\mathcal{T}}$ represents the time-ordering operator. The difference $\Delta_{\infty,t}^{(\textbf{x}\textbf{y})}$ is in trace norm at any finite time $t$ bounded by
\begin{equation}\label{eq:bound_difference}
    \left\| \Delta_{\infty,t}^{(\textbf{x}\textbf{y})} \right\|_{\mathrm{tr}} \le e^{t M} \left\|\Delta_{\infty,0}^{(\textbf{x}\textbf{y})} \right\|_{\mathrm{tr}}+ t e^{tM}  \left\| \mathcal{A}_s  \right\|_{\mathrm{tr}}\, ,
\end{equation}
with
$$
\|\mathcal{M}_t[\cdot]\|_\mathrm{tr} \leq \|\mathcal{L}_1\left[ \cdot\right]\|_\mathrm{tr} + \int {\rm d}\textbf{z} \frac{1}{\|\textbf{x}-\textbf{z}\|_\Gamma^\alpha} \left\|\mathrm{tr}_{\textbf{z}} \circ \mathcal{L}_{12}\circ\left[ \cdot \otimes \tilde{\varrho}_{\infty,t}^{(\textbf{z})}\right] \right\|_\mathrm{tr} \leq \left(E + \int {\rm d}\textbf{z} \frac{1}{\|\textbf{x}-\textbf{z}\|_\Gamma^\alpha} C \right) \|\cdot\|_\mathrm{tr} = M \|\cdot\|_\mathrm{tr} \, ,
$$
where 
$$
    E= 2\|h\|+2\sum_\mu  |\kappa^\mu| ||L^{(k)}_\mu|| ||L_\mu^{(k)\,\dagger}||\, ,
$$
and $C$ as in Lemma~\ref{lemma1}. By assumption, the initial state is continuous, so that  $\|\Delta_{\infty,0}^{(\textbf{x}\textbf{y})} \|_{\mathrm{tr}}$ can be made arbitrarily small by taking $\|\textbf{x} - \textbf{y}\|$ small enough.
Furthermore, the quantity $\left\| \mathcal{A}_s  \right\|_{\mathrm{tr}}$ is  bounded by a continuous function as 
$$
\left\| \mathcal{A}_s  \right\|_{\mathrm{tr}} \le \left|\int {\rm d}\textbf{z} \frac{1}{\|\textbf{x}-\textbf{z}\|_\Gamma^\alpha} -\int {\rm d}\textbf{z} \frac{1}{\|\textbf{y}-\textbf{z}\|_\Gamma^\alpha}\right| C\, .
$$
As such by choosing $\|{\bf x} - {\bf y}\|$ to be small enough,  one can make the whole term in Eq.~\eqref{eq:bound_difference} arbitrarily small, implying continuity of $\tilde{\varrho}_{\infty,t}^{(\textbf{x})}$. \qed

\newpage

\section{IV. Additional results for the long-range dissipative quench}
\begin{figure}[h!]
    \centering
    \includegraphics[width=0.8\textwidth]{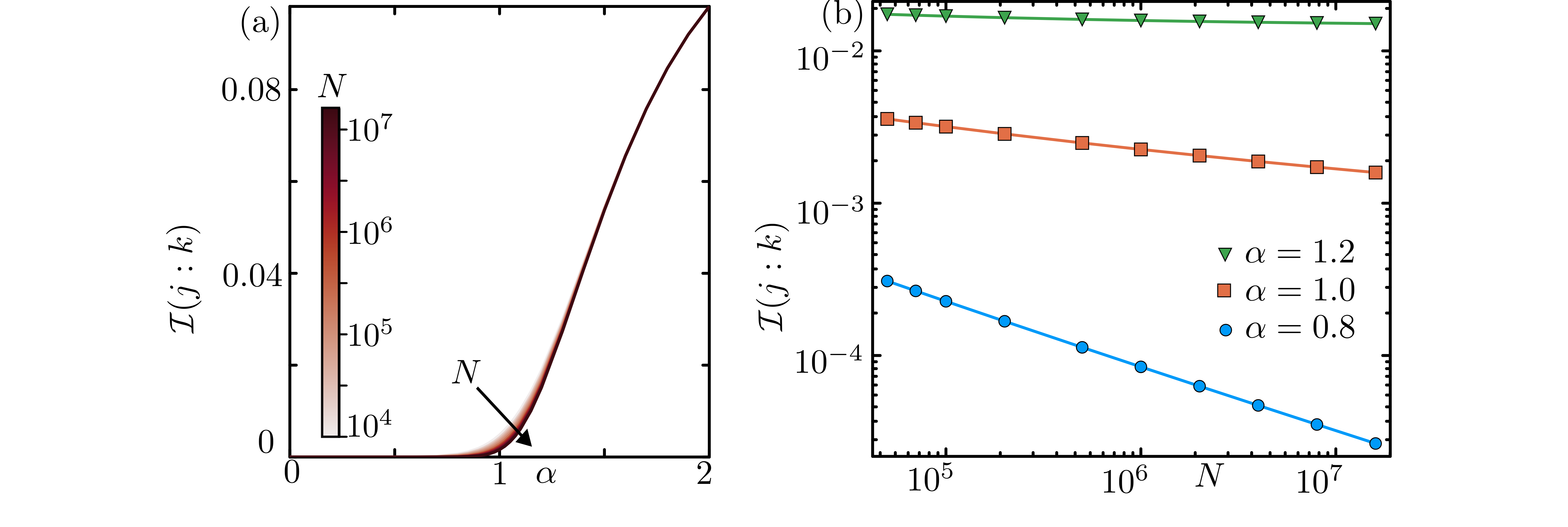}
    \caption{{\bf Quantum mutual information.} We investigate correlations between the sites $j$ and $k$ of the lattice by means of the quantum mutual information $\mathcal{I}(j:k)$. We consider here $j=N/2$, $k=N/2+1$,  $\kappa=0.1\Omega$ and $\Omega t=0.5$. (a) Quantum mutual information $\mathcal{I}(j:k)$ as a function of the interaction range $\alpha$, for different system sizes $N=10^4,\dots16\times10^6$. (b) Log-log plot of the quantum mutual information $\mathcal{I}(j:k)$ as a function of the system size $N$ for $\alpha=0.8,1,1.2$.}
    \label{FigSM}
\end{figure}
In this section, we provide details on the derivation of the collision-model circuit [cf.~Fig.~\ref{Fig2}(b)] we have used to numerically simulate the long-range dissipative quench.
Furthermore, we show additional numerical results on the quantum mutual information between reduced single-particle states.

Since the Hamiltonian and the dissipator of the considered system commute, the evolution can be decomposed into the separate action of a unitary  and a dissipative map, as  $\varrho_{N,t} = e^{t\mathcal{L}_N}[\varrho_{N}] = e^{t\mathcal{D}_N}\circ e^{-i t[H_N,\cdot]}[\varrho_{N}]$. In order to verify the decay of correlations and the validity of Theorem~\ref{theorem}, we are interested in quantifying measures of correlations between two sites, $j$ and $k$, of the one-dimensional system. The reduced state containing this information is defined as $\varrho_{N,t}^{(\{j,k\})} = \tr_{\{j,k\}'}\left(\varrho_{N,t}\right)$, where all particles,  but particles $j$ and $k$, are traced out. With the previously discussed decomposition of the dynamics, the reduced state is given by

\begin{align}\label{eq:collision_model}
  \varrho_{N,t}^{(\{j,k\})} = \left(e^{t\mathcal{L}_{1}^{}}\otimes e^{t\mathcal{
  L}_{1}^{}}\right)\tr_{\{j,k\}'}\left(\prod_{\substack{\ell, m=1\\m<\ell}}^N \mathcal{U}_{\ell m}\right)\left[ \varrho_{N}\right]\, , \quad \mathrm{with} \quad \mathcal{U}_{\ell m}[\cdot] = e^{\frac{-it\sigma_x^{(\ell)}\sigma_x^{(m)}}{c_\alpha^N|\ell-m|^\alpha}}\cdot e^{\frac{it\sigma_x^{(\ell)}\sigma_x^{(m)}}{c_\alpha^N|\ell - m|^\alpha}}\, .
\end{align}

\noindent Here, we have used that the dissipator is a sum of local terms, given by the single-site dissipative maps $\mathcal{L}_{1}^{(j)}$ which act independently on each site. Furthermore, we have used that the single-site dissipator defines a completely positive and trace-preserving map, so that after the partial trace the only remaining dissipative channels are the ones acting on particles $j$ and $k$. For the same reason, all unitary maps $\mathcal{U}_{\ell m}$ with both $\ell,m\neq k,j$ become irrelevant after the partial trace. Since the system is initially in a product state $\varrho_N=\bigotimes_{j\in\Gamma}\varrho_N^{(j)}$ we further  find that
\begin{equation}\label{eq:collision_circuit}
  \varrho_{N,t}^{(\{j,k\})} = \left(e^{t\mathcal{L}_{1}^{}}\otimes e^{t\mathcal{
  L}_{1}^{}}\right)\circ\tr_{N}\left(\mathcal{U}_{k,N}\circ\mathcal{U}_{j,N}\dots \tr_{1}\left(\mathcal{U}_{k,1}\circ\mathcal{U}_{j,1}\left(\mathcal{U}_{j,k}\left[ \varrho_{N}^{(j)}\otimes\varrho_{N}^{(k)}\right]\right)\otimes\varrho_{N}^{(1)}\right)\dots \otimes\varrho_{N}^{(N)}\right)\, . 
\end{equation}
\noindent This formula shows that we can calculate the evolution of the reduced state $\varrho_{N,t}^{(\{j,k\})}$ as follows. First, we consider the interaction $\mathcal{U}_{jk}$ between the particles of interest. Then, we make both site $j$ and site $k$ sequentially interact with one particle in $\{j,k\}'$ and tracing the latter out. After all particles in the remainder of the system have been considered, we can apply the dissipative contributions. This algorithm, schematically illustrated in Fig.~\eqref{Fig2}(b), allows us to investigate the many-body system with a complexity which is equivalent to that of a three-qubit system subject to order $N$ maps, which involve at most two qubits. In the numerical calculations, we have used the initial state $
\varrho_{N}^{(j)} =\left(\begin{array}{cc}
        0.2&  0.2-0.2i\\
        0.2+0.2i & 0.8
    \end{array}\right)\, ,
$
for each site.
In Fig.~\ref{FigSM}(a,b), we show results for the quantum mutual information between the two sites, $\mathcal{I}(j:k)= \mathcal{S}(\varrho_{N,t}^{(\{j\})}) + \mathcal{S}(\varrho_{N,t}^{(\{k\})}) - \mathcal{S}(\varrho_{N,t}^{(\{j,k\})})$, with $\mathcal{S}(\varrho) = -\tr\left(\varrho\log(\varrho)\right)$ being the von Neumann entropy.
We see that the quantum mutual information decreases with the system size in the strong long-range regime $\alpha\le1$. Its behavior as a function of $N$ [shown in Fig.~\ref{FigSM}(b)] is in principle not the same shown in the inset of Fig.~\ref{Fig3}(b) by the correlations $C_{zz}$, due to the definition of the quantum mutual information through  entropy functions.

\section{V. Extension to quasi-local operators}
\noindent Beyond statements about expectation values of local observables,  Theorem~1 immediately  applies also to non-strictly local observables and correlation functions.\\
To understand this, we can indeed consider the so-called ``quasi-local algebra" of a many-body system~\cite{Bratelli1981}, which  contains operators that are not strictly local but are still convergent in the norm topology. These quasi-local operators $A$ can be approximated, with arbitrary precision, by strictly local operators $A_\epsilon$. This implies that we can use our results to show the noninteracting character of the many-body state also on operators which are almost localized yet extended to the whole many-body system. This is because we can always find a suitable local operator $A_\epsilon$ such that 
$
    |\langle A_\epsilon \rangle  - \langle A  \rangle| < \epsilon \, ,
$
with $\epsilon>0$ small at pleasure, and given the fact that the expectation on local operators $\langle A_\epsilon \rangle$ shows a noninteracting character. 
\section{VI. Spontaneous symmetry breaking on the level of the emergent locally noninteracting dynamics}
\noindent First of all, we note that in our approach, we first take the limit $N\to\infty$ which gives rise to the mean-field equations of motion for the systems. Starting from a dynamical (Lindblad) generator obeying a given symmetry, one also has that the mean-field equations obey the same symmetry of the generator. In general, the mean-field equations of motion have a unique stable solution which also obeys the same symmetry of the generator. Spontaneous symmetry breaking in these long-range interacting systems occurs when the mean-field equations feature two, or more, stable solutions which break the symmetry of the generator (see e.g. Ref.~\cite{Boneberg2022}). Each of these solutions is dynamically approached according to which initial state is chosen for the system.

\noindent Importantly, we note that mean-field dynamics are typically valid only when the initial state of the system has low correlations (in our case we considered  initial product states). These states are usually not symmetric and they thus approach only one of the different stable solutions available in the symmetry-broken phase. The way to restore the symmetry in this setup is to consider the state given by the sum of different (product) mean-field solutions, associated with different initial states, such that the incoherent mixture state obeys the symmetry of the generator. This leads to a stationary state that is given by the sum of all stable (product) solutions and is thus symmetric. In the complementary approach, in which one is able to first take the limit $t\to\infty$, the stationary state of the system  is symmetric since for any finite $N$ there is no symmetry breaking. Spontaneous symmetry breaking then solely occurs when taking the limit $N\to\infty$, manifesting through the emergence of an incoherent mixture state which, at the level of local observables,  behaves as the incoherent sum over the (product) stable solutions of the mean-field equations. \\

\noindent To concretely understand the above mechanism, let us consider the simple mean-field equation of the magnetization $m$ in the classical Ising model. The dimensionless version of the equation is given by
\begin{equation}
    \dot{m}=-m+\tanh(\beta m)\, ,
\end{equation}
where $\beta$ is a dimensionless inverse temperature. The above equation is invariant for $m\to -m$, which constitutes the $Z_2$ symmetry of the model. For sufficiently small values of $\beta$ the system features a unique solution $m=0$. For $\beta$ large enough, we instead have that two possible solutions $m_1$, $m_2$ exist, with opposite magnetization $m_1=-M$ and $m_2=M$. In this symmetry-broken regime, one can see that initial states with negative magnetization will approach solution $m_1$, while those with positive magnetization will approach $m_2$. To construct a symmetric state, one can consider the sum of an initial (product) state with $m=-\bar{m}$ plus another initial (product) state with $m=\bar{m}$. The latter incoherent initial state converges to a stationary state given by the sum of the (product) state with magnetization $-M$ and the (product) state with magnetization $M$. Note that this is different from having a stationary state made by a single (product) state with magnetization equal to zero.
\end{document}